\documentclass{scrartcl}
\KOMAoptions{paper=letter}

\usepackage[todo,fixcref]{group}
\usepackage{mathtools, thmtools, nicefrac}
\usepackage{bbm}
\usepackage{thm-restate}
\usepackage{multirow}

\newcommand{\NAE}{\mathrm{NAE}}
\newcommand{\D}{\mathcal D}
\newcommand{\T}[1]{t_{#1}}
\newcommand{\Ri}[1]{\mathrel{R_{#1}}}

\colorlet{cola}{Set1-A} \colorlet{colb}{Set1-B}
\colorlet{colc}{Set1-C}
\colorlet{cold}{Set1-D}
\colorlet{cole}{Set1-E}
\definecolor{ink}{HTML}{20363B}\colorlet{ordfill}{black!7}
\colorlet{accent}{colc}
\colorlet{Xc}{cola}\colorlet{Yc}{colb}\colorlet{Zc}{colc}
\colorlet{Adraw}{colb}\colorlet{Afill}{colb!15}
\colorlet{Bdraw}{cola}\colorlet{Bfill}{cola!15}
\colorlet{Cone}{colb}\colorlet{Ctwo}{colc}
\colorlet{sepdraw}{cole}\colorlet{sepfill}{cole!25}

\crefname{observation}{Observation}{Observations}

\title{Recognizing $k$-inducible Digraphs is NP-complete}
\author{Felix Brandt \qquad Fabian Frank\\
Technical University of Munich}

\begin{document}
\maketitle

\begin{abstract}
A digraph is \emph{$k$-inducible} if its arcs correspond to the pairwise majority relation of a preference profile with $k$ voters. By McGarvey's theorem, every asymmetric digraph is $k$-inducible for some~$k$, but for every fixed~$k$ there are digraphs that are not $k$-inducible. 
Whether a digraph is $1$-inducible or $2$-inducible can easily be decided in polynomial time. 
The complexity of deciding whether a digraph is $k$-inducible for every fixed $k\geq 3$ has been a long-standing open problem. We prove that deciding whether a tournament is $3$-inducible is NP-complete and
then leverage this result to show that deciding whether a digraph is $k$-inducible is NP-complete for every fixed $k\ge3$. 
Our results imply that, unless $\mathrm{P}=\mathrm{NP}$, $k$-inducible digraphs cannot be characterized by finitely many forbidden induced subdigraphs.
\end{abstract}

\section{Introduction} \label{sec:intro}

Given a list of linear orders over some set of alternatives, interpreted as preferences of voters, the \emph{pairwise majority relation} of these orders forms an asymmetric digraph. When the number of voters is odd, this digraph is a tournament. A classic theorem by \citet{McGa53a} states that every asymmetric digraph arises in this way, provided sufficiently many voters are available. For digraphs on $n$ alternatives, the number of voters required is of order $\Theta(n/\log n)$ \citep{Stea59a,ErMo64a}. Hence, for every fixed~$k$, only a restricted class of digraphs can be induced by $k$~voters. Following \citet{BBG+14a}, we call a digraph \emph{$k$-inducible} if it is the majority relation of some preference profile with $k$~voters, and we call the smallest such~$k$ its \emph{majority dimension}. Inducibility by few voters has been studied in combinatorics \citep{ABKK+06a,MSW11a} and in computational social choice \citep{BCE+14a}.

For one or two voters, inducibility is well understood. A digraph is $1$-inducible if and only if it is a transitive tournament. The majority relation of two voters coincides with their Pareto relation, so the $2$-inducible digraphs are exactly the partial orders of dimension at most~$2$, which were characterized by \citet{DuMi41a} and can be recognized in polynomial time \citep{Yann82a}. Three voters are thus the first non-trivial case. Every tournament on at most seven alternatives is $3$-inducible \citep{BBG+14a}, but there are tournaments on eight alternatives that are not \citep{ShTo09a}. For five voters, the corresponding threshold has only recently been narrowed down: every tournament on at most 12 alternatives is $5$-inducible, but there are tournaments on 22 alternatives that are not \citep{ChHa26a}. \citet{ChHa26a} also showed that the quadratic residue tournament on 23 alternatives
is not $5$-inducible. \citet{BBG+14a} characterized $3$-inducible tournaments as those whose arc set can be partitioned into a transitive part and an acyclic, transitively reorientable part, but noted that there is no obvious efficient way to check this condition. A more symmetric characterization is the following: a tournament is $3$-inducible if and only if its arc set can be partitioned into three sets such that the union of any two of them is acyclic. Equivalently, its arcs can be colored with three colors such that every directed cycle contains arcs of all three colors, or it has three pairwise disjoint feedback arc sets.
Recently, there has been some progress on hardness results related to $3$-inducible tournaments: \citet{Pete26a} and \citet{Mada26a} have shown that computing Kemeny rankings is NP-hard, even when there are only three voters.

The complexity of \textsc{Check-$k$-Majority}, the problem of deciding whether a given digraph is $k$-inducible, has been open for every fixed $k\ge 3$ at least since \citet{BHKS12a}.
The analogous question for the Pareto relation, rather than the majority relation, was answered long ago. The \emph{dimension} of a partial order is the smallest number of linear orders whose intersection it is, i.e., the smallest number of voters whose Pareto relation it is \citep{DuMi41a}. \citet{Yann82a} showed that deciding whether a partial order has dimension at most~$k$ is NP-complete for every fixed $k\ge3$. For majority rule,
the corresponding question has remained open. We resolve it for every fixed $k\ge3$. Our main result concerns three voters.

\begin{restatable}{theorem}{mainthm}\label{thm:main}
Deciding whether a tournament is $3$-inducible is NP-complete.
\end{restatable}

Membership in NP is immediate since three linear orders serve as a certificate. \Cref{thm:main} extends to every fixed number of voters.

\begin{restatable}{theorem}{morevotersthm}\label{thm:more-voters}
For every fixed $k\ge3$, deciding whether a digraph is $k$-inducible is NP-complete.
\end{restatable}

For odd~$k$, only tournaments are $k$-inducible, so the problem remains NP-complete for tournaments. We prove \Cref{thm:more-voters} in \Cref{app:more-voters}. The reduction pads a $3$-voter profile with voters whose preferences are fixed up to reversal and uses Ramsey's theorem to enforce this shape. It is polynomial for every fixed~$k$, but of astronomical size.

These results have several consequences.
First, computing the majority dimension of a tournament is NP-hard because a tournament is $3$-inducible if and only if its majority dimension is at most~$3$. Second, $k$-inducibility is preserved under taking induced subdigraphs. Unless $\mathrm{P}=\mathrm{NP}$, it therefore cannot be characterized by a finite set of forbidden induced subdigraphs for any fixed $k\ge3$. For $k=3$, this applies to the forbidden subtournaments found by \citet{ShTo09a}, and the characterization by \citet{BBG+14a} does not yield an efficient algorithm.

Our proof of \Cref{thm:main} is a reduction from monotone not-all-equal $3$-satisfiability or, equivalently, $2$-colorability of $3$-uniform hypergraphs \citep{Lova73a,Scha78a,GaJo79a}. 
The three voters play different roles. A rigid gadget consisting of six \emph{anchors} and a sequence of \emph{separators} forces one of the voters to rank the remaining alternatives according to a prescribed ordered partition. This part of the construction works for arbitrary tournaments and ordered partitions. Since the ranking of this voter is essentially fixed, it acts as a predictable tie-breaker on all pairs on which the other two voters disagree. These two voters encode a truth assignment: the value of a variable is determined by which of them ranks all occurrences of the variable before a threshold alternative~$z$. A clause gadget on seven alternatives can be induced with the third voter following the prescribed order if and only if the three variables of the clause do not all have the same value. For an instance with $m$ clauses, the resulting tournament has $10m+7$ alternatives. The proof is elementary and self-contained. It only relies on simple properties of three-voter profiles, such as that at most one voter disagrees with each arc and that every voter disagrees with exactly one arc of every directed triangle.

The remainder of the paper is structured as follows. \Cref{sec:preliminaries} introduces our notation. \Cref{sec:pinning} shows how to force one voter to respect a prescribed ordered partition. \Cref{sec:tournament} defines the problem we reduce from and presents the reduction: \Cref{sec:construction} constructs the tournament, and \Cref{sec:correctness} analyzes the clause and variable gadgets and proves \Cref{thm:main}. \Cref{app:more-voters} extends the result to more voters.

\section{Preliminaries} \label{sec:preliminaries}\label{sec:majority}

A \emph{tournament} $T=(V,E)$ consists of a finite set $V$ of alternatives and an asymmetric and complete relation $E\subseteq V\times V$, i.e., for all distinct $y,y'\in V$, exactly one of $(y,y')$ and $(y',y)$ is in~$E$. We write $y\to y'$ if $(y,y')\in E$ and call $y\to y'$ an \emph{arc}. For $V'\subseteq V$, the tournament $(V',E\cap(V'\times V'))$ is the \emph{subtournament} of $T$ induced by~$V'$.

Let $N=\{1,2,3\}$ be a set of three voters. The preferences of each voter~$i$ are given as a linear order $R_i$ over~$V$, and we write $y\Ri{i}y'$ if voter~$i$ prefers $y$ to~$y'$. A \emph{preference profile} $R=(R_1,R_2,R_3)$ associates preferences with each voter. When we specify the preferences of a voter explicitly, we list the alternatives from most to least preferred. The profile $R$ \emph{induces} the tournament $T$ if $T$ is the majority digraph of~$R$, i.e., if for every arc $y\to y'$ at least two of the three voters prefer $y$ to~$y'$. A tournament is \emph{$3$-inducible} if it is induced by some $3$-voter profile. Restricting a profile that induces $T$ to a subset of the alternatives yields a profile that induces the corresponding subtournament. 

Voter $i$ \emph{disagrees} with the arc $y\to y'$ if $y'\Ri{i}y$, and \emph{agrees} with it otherwise. Our hardness proof relies on the following three simple properties of $3$-voter profiles.

\begin{observation}\label{obs:counting}
In every $3$-voter profile that induces a tournament:
\begin{enumerate}[leftmargin=3.2em]
\item[(D1)] \textbf{at most one} voter disagrees with each arc; so if one voter disagrees with an arc, the other two agree with it;
\item[(D2)] on a directed triangle, every voter disagrees with \textbf{exactly} one arc, and every arc has exactly one disagreeing voter;
\item[(D3)] if $y\to y'$, $y'\to y''$, and $y\to y''$, then every voter that disagrees with $y\to y''$ also disagrees with $y\to y'$ or with $y'\to y''$.
\end{enumerate}
\end{observation}

\begin{proof}
(D1) holds because every arc is supported by at least two of the three voters. For (D2), note that no voter agrees with all three arcs of a directed triangle, so each of the three voters disagrees with at least one of them. By (D1), each of the three arcs has at most one disagreeing voter. Hence, each voter disagrees with exactly one arc, and each arc has exactly one disagreeing voter. For (D3), suppose that a voter disagrees with $y\to y''$ but agrees with $y\to y'$ and $y'\to y''$. Then it prefers $y$ to~$y'$, $y'$ to~$y''$, and $y''$ to~$y$, contradicting the transitivity of its preferences.
\end{proof}

Finally, let $\D=(D_1,\dots,D_K)$ be an ordered partition of a set of alternatives into \emph{blocks}. A voter \emph{respects} $\D$ if it prefers every alternative in $D_k$ to every alternative in $D_l$ whenever $k<l$; the order within a block is arbitrary.

\section{Restricting the Preferences of One Voter} \label{sec:pinning}

In this section, we describe how to enforce a fixed structure on the preferences of one of the voters. Since the majority relation does not depend on the names of the voters, we cannot require a specific voter to have this structure. Nevertheless, we can identify a voter with the following method. By (D2), every voter disagrees with exactly one arc of a directed triangle, so a fixed arc of a directed triangle singles out one voter. We use this method throughout this section.

In the following, let $T_0=(V,E_0)$ be a tournament with $V\neq\emptyset$, and let $\D=(D_1,\dots,D_K)$ be an ordered partition of~$V$. For $y\in V$, we write $\mathit{ind}(y)$ for the index of the block containing~$y$. Based on $T_0$, we construct a tournament $T_0^{\D}$ with $K+5$ new alternatives and the following properties:
\begin{itemize}
\item every $3$-voter profile that induces $T_0^{\D}$ contains a voter that respects~$\D$;
\item every $3$-voter profile that induces $T_0$ and contains a voter that respects~$\D$ can be extended to a profile that induces~$T_0^{\D}$.
\end{itemize}

\subsection{The Tournaments \texorpdfstring{$H_+$ and $H_-$}{H+ and H-}} \label{sec:template}

\begin{figure}[tbp]
\centering
\begin{tikzpicture}[y=0.72cm,x=0.95cm,
  v/.style={circle,draw=ink!70,line width=.5pt,minimum size=7mm,inner sep=0pt,font=\small},
  anc/.style={v,fill=ordfill},
  org/.style={v,fill=white},
  sepv/.style={v,fill=sepfill,draw=sepdraw},
  arc/.style={-Latex,draw=ink!65,line width=.55pt},
  hi/.style={-Latex,draw=accent,line width=1.3pt}
]
\node[anc]  (t2) at ( 2.0,7) {$\T2$};
\node[anc]  (t3) at (-0.5,6) {$\T3$};
\node[anc]  (t4) at ( 3.6,5) {$\T4$};
\node[org]  (t0) at (-2.5,4) {$\T0$};
\node[sepv] (t1) at (-1.3,3) {$\T1$};
\node[anc]  (t5) at ( 3.6,2) {$\T5$};
\node[anc]  (t6) at ( 0.2,1) {$\T6$};
\node[anc]  (t7) at ( 2.0,0) {$\T7$};
\draw[arc] (t0) to[out=85,in=180] (t2);
\draw[arc] (t7) to[out=180,in=-85] (t0);
\draw[arc] (t7) -- (t2);
\draw[arc] (t7) -- (t3);
\draw[arc] (t6) -- (t2);
\draw[arc] (t6) -- (t3);
\draw[arc] (t6) -- (t1);
\draw[arc] (t1) -- (t3);
\draw[hi] (t5) -- (t4);
\draw[hi] (t0) -- (t1);
\end{tikzpicture}
\caption{The tournament $H_+$. The heights follow the preferences $R^+_3$ of voter~$3$, so $\T2$ is at the top and $\T7$ at the bottom. Arcs that are not drawn point downward; the drawn gray arcs point upward and are exactly the arcs with which voter~$3$ disagrees. Highlighted are the arc $\T5\to\T4$, with which voter~$3$ disagrees, and the arc $\T0\to\T1$, with which it agrees. The tournament $H_-$ differs only in the direction of the arc between $\T0$ and~$\T1$. In the extension of \Cref{sec:extension}, the gray alternatives become the anchors, $\T0$ is replaced by an original alternative, and $\T1$ by a separator.}
\label{fig:hplus}
\end{figure}
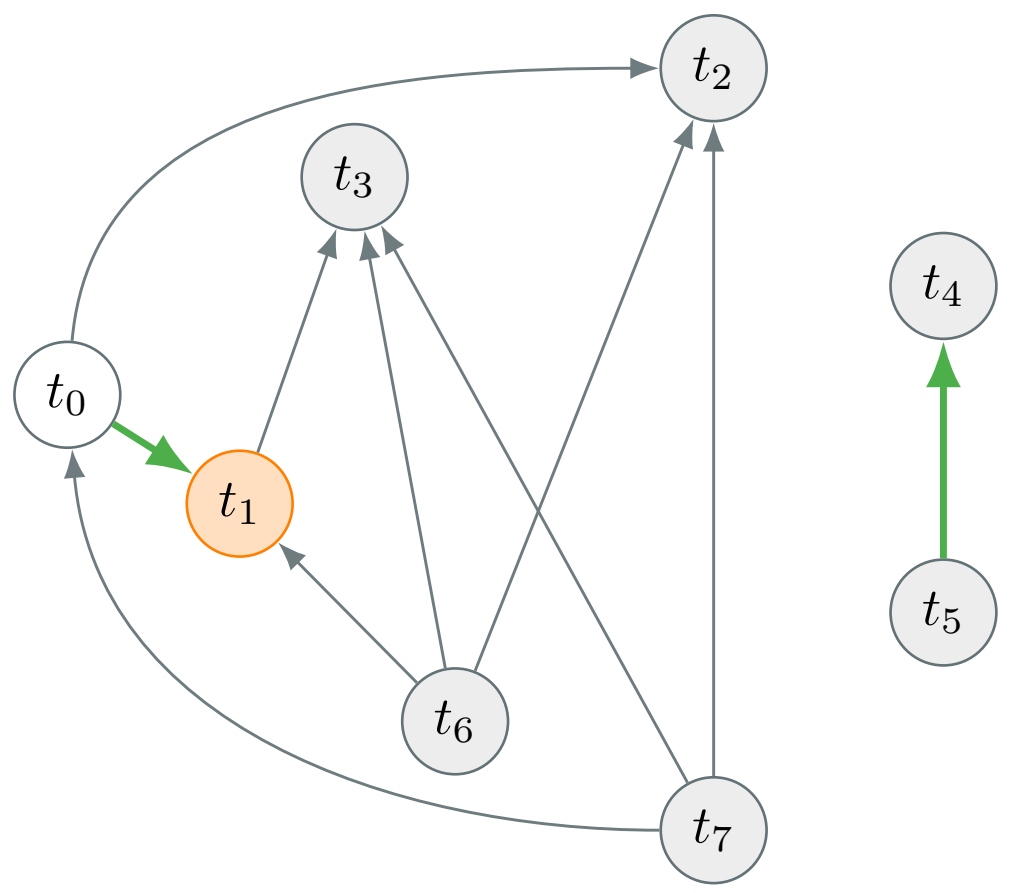

The main idea behind the construction of $T_0^{\D}$ lies in the properties of the following two tournaments $H_+$ and~$H_-$. Consider the profile $R^+=(R^+_1,R^+_2,R^+_3)$ given by
\[
R^+_1\colon \T7\,\T0\,\T6\,\T2\,\T1\,\T3\,\T5\,\T4,\qquad
R^+_2\colon \T5\,\T4\,\T6\,\T1\,\T7\,\T3\,\T0\,\T2,\qquad
R^+_3\colon \T2\,\T3\,\T4\,\T0\,\T1\,\T5\,\T6\,\T7,
\]
and let $H_+$ be the tournament it induces (\Cref{fig:hplus}). Voters $1$ and~$2$ order $\T0$ and $\T1$ differently, so the arc between them is decided by voter~$3$. Since $\T0$ and $\T1$ are adjacent in $R^+_3$, exchanging them there reverses this arc and leaves all other arcs unchanged; we write $R^-$ for the resulting profile and $H_-$ for the tournament it induces. Thus, $H_+$ contains $\T0\to\T1$, $H_-$ contains $\T1\to\T0$, and the two tournaments agree on all other pairs.

Observe that in both profiles, voter~$3$ is the only voter that disagrees with the arc $\T5\to\T4$, and it agrees with the arc between $\T0$ and~$\T1$ (as does one of the other two voters). The next lemma shows that this is forced: in every profile that induces $H_+$ or~$H_-$, the unique voter that disagrees with $\T5\to\T4$ agrees with the arc between $\T0$ and~$\T1$.

\begin{lemma}\label{lem:template}
If a $3$-voter profile induces $H_+$ or~$H_-$, then exactly one voter disagrees with the arc $\T5\to\T4$, and this voter agrees with the arc between $\T0$ and~$\T1$.
\end{lemma}

\begin{figure}[tbp]
\centering
\resizebox{\linewidth}{!}{\begin{tikzpicture}[
  tv/.style={circle,draw=ink,line width=.5pt,minimum size=6.6mm,inner sep=0pt,font=\footnotesize,fill=white},
  tarc/.style={-Latex,line width=.55pt,draw=ink!70},
  key/.style={line width=1.3pt,draw=ink},
  bad/.style={draw=ink,line width=1.5pt},
  lab/.style={circle,inner sep=0pt,minimum size=4.2mm,font=\scriptsize,line width=.8pt},
  tcap/.style={font=\footnotesize,anchor=north}
]
\begin{scope}[shift={(0.00,0.00)}]
\node[tv] (n0v0) at (-1.45,-1.15) {$\T0$};
\node[tv] (n0v1) at (0.00,1.25) {$\T1$};
\node[tv] (n0v3) at (1.45,-1.15) {$\T3$};
\draw[tarc] (n0v0) -- node[lab,draw=Xc,text=Xc,fill=white] {$X$} (n0v1);
\draw[tarc] (n0v1) -- node[lab,draw=Zc,fill=Zc!20,text=black] {$Z$} (n0v3);
\draw[tarc] (n0v3) -- node[lab,draw=Yc,fill=Yc!20,text=black] {$Y$} (n0v0);
\node[tcap] at (0,-1.72) {(1) name the other voters $Y$, $Z$};
\end{scope}
\begin{scope}[shift={(5.30,0.00)}]
\node[tv] (n1v4) at (0.00,1.25) {$\T4$};
\node[tv] (n1v1) at (0.00,-1.25) {$\T1$};
\node[tv] (n1v5) at (-1.55,0.00) {$\T5$};
\node[tv] (n1v3) at (1.55,0.00) {$\T3$};
\draw[tarc] (n1v5) -- node[lab,draw=Xc,text=Xc,fill=white] {$X$} (n1v4);
\draw[tarc] (n1v1) -- node[lab,draw=Zc,text=Zc,fill=white] {$Z$} (n1v3);
\draw[tarc,key] (n1v4) -- node[lab,draw=Yc,fill=Yc!20,text=black] {$Y$} (n1v1);
\draw[tarc] (n1v1) -- node[lab,draw=Zc,fill=Zc!20,text=black] {$Z$} (n1v5);
\draw[tarc] (n1v3) -- node[lab,draw=Xc,fill=Xc!20,text=black] {$X$} (n1v4);
\node[tcap] at (0,-1.72) {(2) $\T4\to\T1$ avoids $X$ and~$Z$};
\end{scope}
\begin{scope}[shift={(10.60,0.00)}]
\node[tv] (n2v6) at (0.00,1.25) {$\T6$};
\node[tv] (n2v3) at (0.00,-1.25) {$\T3$};
\node[tv] (n2v0) at (-1.55,0.00) {$\T0$};
\node[tv] (n2v4) at (1.55,0.00) {$\T4$};
\draw[tarc] (n2v3) -- node[lab,draw=Yc,text=Yc,fill=white] {$Y$} (n2v0);
\draw[tarc] (n2v3) -- node[lab,draw=Xc,text=Xc,fill=white] {$X$} (n2v4);
\draw[tarc,key] (n2v6) -- node[lab,draw=Zc,fill=Zc!20,text=black] {$Z$} (n2v3);
\draw[tarc] (n2v0) -- node[lab,draw=Xc,fill=Xc!20,text=black] {$X$} (n2v6);
\draw[tarc] (n2v4) -- node[lab,draw=Yc,fill=Yc!20,text=black] {$Y$} (n2v6);
\node[tcap] at (0,-1.72) {(3) $\T6\to\T3$ avoids $Y$ and~$X$};
\end{scope}
\begin{scope}[shift={(0.00,-3.85)}]
\node[tv] (n3v4) at (-1.45,-1.15) {$\T4$};
\node[tv] (n3v0) at (0.00,1.25) {$\T0$};
\node[tv] (n3v1) at (1.45,-1.15) {$\T1$};
\draw[tarc] (n3v4) -- node[lab,draw=Yc,fill=Yc!20,text=black] {$Y$} (n3v0);
\draw[tarc] (n3v0) -- node[lab,draw=Xc,text=Xc,fill=white] {$X$} (n3v1);
\draw[tarc] (n3v4) -- node[lab,draw=Yc,text=Yc,fill=white] {$Y$} (n3v1);
\node[tcap] at (0,-1.72) {(4) (D3) on $\T4\to\T0\to\T1$};
\end{scope}
\begin{scope}[shift={(5.30,-3.85)}]
\node[tv] (n4v0) at (-1.45,-1.15) {$\T0$};
\node[tv] (n4v5) at (0.00,1.25) {$\T5$};
\node[tv] (n4v4) at (1.45,-1.15) {$\T4$};
\draw[tarc] (n4v0) -- node[lab,draw=Zc,fill=Zc!20,text=black] {$Z$} (n4v5);
\draw[tarc] (n4v5) -- node[lab,draw=Xc,text=Xc,fill=white] {$X$} (n4v4);
\draw[tarc] (n4v4) -- node[lab,draw=Yc,text=Yc,fill=white] {$Y$} (n4v0);
\node[tcap] at (0,-1.72) {(5) (D2) on $\T0\to\T5\to\T4$};
\end{scope}
\begin{scope}[shift={(10.60,-3.85)}]
\node[tv] (n5v0) at (-1.45,-1.15) {$\T0$};
\node[tv] (n5v5) at (0.00,1.25) {$\T5$};
\node[tv] (n5v6) at (1.45,-1.15) {$\T6$};
\draw[tarc] (n5v0) -- node[lab,draw=Zc,text=Zc,fill=white] {$Z$} (n5v5);
\draw[tarc] (n5v5) -- node[lab,draw=Xc,fill=Xc!20,text=black] {$X$} (n5v6);
\draw[tarc] (n5v0) -- node[lab,draw=Xc,text=Xc,fill=white] {$X$} (n5v6);
\node[tcap] at (0,-1.72) {(6) (D3) on $\T0\to\T5\to\T6$};
\end{scope}
\begin{scope}[shift={(0.00,-7.70)}]
\node[tv] (n6v6) at (0.00,1.25) {$\T6$};
\node[tv] (n6v2) at (0.00,-1.25) {$\T2$};
\node[tv] (n6v4) at (-1.55,0.00) {$\T4$};
\node[tv] (n6v5) at (1.55,0.00) {$\T5$};
\draw[tarc] (n6v4) -- node[lab,draw=Yc,text=Yc,fill=white] {$Y$} (n6v6);
\draw[tarc] (n6v5) -- node[lab,draw=Xc,text=Xc,fill=white] {$X$} (n6v6);
\draw[tarc,key] (n6v6) -- node[lab,draw=Zc,fill=Zc!20,text=black] {$Z$} (n6v2);
\draw[tarc] (n6v2) -- node[lab,draw=Xc,fill=Xc!20,text=black] {$X$} (n6v4);
\draw[tarc] (n6v2) -- node[lab,draw=Yc,fill=Yc!20,text=black] {$Y$} (n6v5);
\node[tcap] at (0,-1.72) {(7) $\T6\to\T2$ avoids $Y$ and~$X$};
\end{scope}
\begin{scope}[shift={(5.30,-7.70)}]
\node[tv] (n7v7) at (0.00,1.25) {$\T7$};
\node[tv] (n7v0) at (0.00,-1.25) {$\T0$};
\node[tv] (n7v5) at (-1.55,0.00) {$\T5$};
\node[tv] (n7v1) at (1.55,0.00) {$\T1$};
\draw[tarc] (n7v0) -- node[lab,draw=Zc,text=Zc,fill=white] {$Z$} (n7v5);
\draw[tarc] (n7v0) -- node[lab,draw=Xc,text=Xc,fill=white] {$X$} (n7v1);
\draw[tarc,key] (n7v7) -- node[lab,draw=Yc,fill=Yc!20,text=black] {$Y$} (n7v0);
\draw[tarc] (n7v5) -- node[lab,draw=Xc,fill=Xc!20,text=black] {$X$} (n7v7);
\draw[tarc] (n7v1) -- node[lab,draw=Zc,fill=Zc!20,text=black] {$Z$} (n7v7);
\node[tcap] at (0,-1.72) {(8) $\T7\to\T0$ avoids $Z$ and~$X$};
\end{scope}
\begin{scope}[shift={(10.60,-7.70)}]
\fill[black!8] (0.00,1.25) -- (0.00,-1.25) -- (1.55,0.00) -- cycle;
\node[tv] (n8v7) at (0.00,1.25) {$\T7$};
\node[tv] (n8v2) at (0.00,-1.25) {$\T2$};
\node[tv] (n8v5) at (-1.55,0.00) {$\T5$};
\node[tv] (n8v1) at (1.55,0.00) {$\T1$};
\draw[tarc] (n8v2) -- node[lab,draw=Yc,text=Yc,fill=white] {$Y$} (n8v5);
\draw[tarc] (n8v5) -- node[lab,draw=Xc,text=Xc,fill=white] {$X$} (n8v7);
\draw[tarc] (n8v1) -- node[lab,bad,fill=white,text=Zc] {$Z$} (n8v7);
\draw[tarc] (n8v7) -- node[lab,bad,fill=Zc!20,text=black] {$Z$} (n8v2);
\draw[tarc] (n8v2) -- (n8v1);
\node[tcap] at (0,-1.72) {(9) two labels $Z$: contradiction};
\end{scope}
\end{tikzpicture}}
\caption{Proof of \Cref{lem:template} for $H_+$. Each arc is labeled with the voter that disagrees with it. Outlined labels are known, from earlier panels or from the assumption that $\T5\to\T4$ and $\T0\to\T1$ have label~$X$; filled labels are derived in the panel. By (D2), the three arcs of a directed triangle have different labels. In panels (2), (3), (7), and~(8), the bold arc lies on two directed triangles, whose known labels leave only one choice for it; the other filled labels follow. In panels (4) and~(6), (D3) passes the label of the long arc on to the short arc whose label is not yet known. In panel~(9), the shaded directed triangle $\T1\to\T7\to\T2\to\T1$ receives the label~$Z$ twice, contradicting (D2).}
\label{fig:template}
\end{figure}
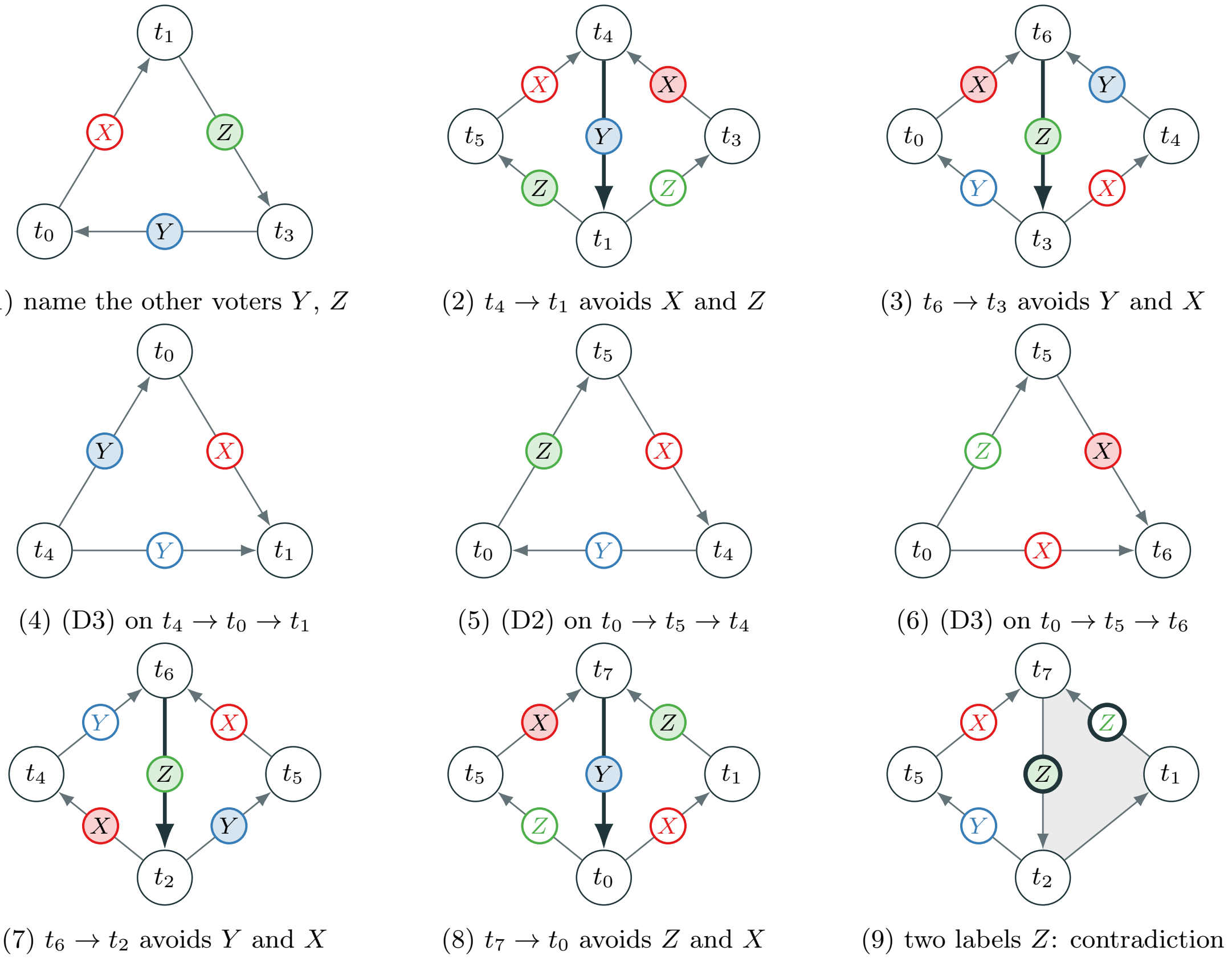

\begin{proof}
Since $\T0\to\T5\to\T4\to\T0$ is a directed triangle, (D2) implies that exactly one voter disagrees with $\T5\to\T4$; call this voter~$X$.

\emph{We first consider $H_+$.} Assume for the sake of contradiction that $X$ disagrees with $\T0\to\T1$. Every arc of $H_+$ lies on a directed triangle; this is easy to check in \Cref{fig:hplus}.\footnote{Every drawn arc $y\to x$ in \Cref{fig:hplus}, together with an alternative $z$ whose height lies between those of $x$ and $y$ and that has no drawn arc to $x$ or~$y$, forms the directed triangle $x\to z\to y\to x$. Such triangles cover all $28$ arcs; for instance, $\T4$ and $\T5$ serve as $z$ for every drawn arc that spans their height. We also verified the claim by computer.} Thus, by (D2), every arc has a unique voter that disagrees with it, and we label every arc of $H_+$ with this voter. By assumption, $\T5\to\T4$ and $\T0\to\T1$ both have label~$X$. Let $Y$ and $Z$ be the labels of $\T3\to\T0$ and $\T1\to\T3$. By (D2), the directed triangle $\T0\to\T1\to\T3\to\T0$ carries all three labels, so $X$, $Y$, and $Z$ are the three voters. We now extend the labeling with the following three moves.
\begin{itemize}
\item If two arcs of a directed triangle have known labels, the third arc receives the remaining label, by (D2).
\item If two directed triangles share an arc, and each of them has another arc with a known label, where the two known labels are different, then the shared arc receives the third label, by (D2); see \Cref{fig:template}(2).
\item If the long arc of a transitive triple and one of its short arcs have known, different labels, then the other short arc receives the label of the long arc, by (D3); see \Cref{fig:template}(4).
\end{itemize}
Applying these moves as shown in panels (1)--(8) of \Cref{fig:template} determines more and more labels. In panel~(9), the arcs $\T1\to\T7$ and $\T7\to\T2$ of the directed triangle $\T1\to\T7\to\T2\to\T1$ both receive the label~$Z$, which contradicts (D2). Thus, our assumption was wrong, and $X$ agrees with $\T0\to\T1$.

\emph{For $H_-$}, we use a symmetry. Reverse the preferences of all three voters and rename $\T i$ as $\T{9-i}$ for $i=2,\dots,7$. This operation is an involution, and it maps $R^+$ to a profile that induces~$H_-$. Consequently, it maps the profiles that induce $H_-$ to profiles that induce~$H_+$. It also maps the arc $\T5\to\T4$ to itself and the arc $\T1\to\T0$ to $\T0\to\T1$, while every voter keeps disagreeing with the same arcs. The claim for $H_-$ therefore follows from the claim for~$H_+$.
\end{proof}

\subsection{The Extension} \label{sec:extension}

The power of \Cref{lem:template} is that whenever a tournament contains $H_+$ or $H_-$ as a subtournament, the voter that disagrees with the arc $\T5\to\T4$ must agree with the arc between $\T0$ and~$\T1$. Thus, we can force a voter to agree with an arc $y\to y'$ by adding six alternatives in the roles of $\T2,\dots,\T7$ and letting $y$ and $y'$ take the roles of $\T0$ and~$\T1$.

The six new alternatives can be reused for many such pairs, but each alternative of $V$ must always play the same role, since $\T0$ and $\T1$ relate differently to $\T2,\dots,\T7$. We therefore let each alternative of $V$ play the role of~$\T0$. For the role of~$\T1$, we create a new alternative $s_k$ for each $k=1,\dots,K-1$; we call these alternatives \emph{separators}. Separator $s_k$ is placed between the blocks $D_k$ and~$D_{k+1}$. If a voter agrees with all arcs between $V$ and the separators, it prefers every block before $s_k$ to every block after~$s_k$, and hence respects~$\D$.

Formally, $T_0^{\D}$ has the alternatives in~$V$, which we call \emph{original}, the separators $s_1$, \dots, $s_{K-1}$, and six \emph{anchors} $\T2$, \dots, $\T7$. The arcs among the original alternatives are those of~$T_0$, and the remaining arcs are as follows.
\begin{enumerate}[leftmargin=3.2em]
\item[(E1)] For each original alternative $y$ and each~$k$, we have $y\to s_k$ if $\mathit{ind}(y)\le k$, and $s_k\to y$ otherwise. Moreover, $s_k\to s_l$ whenever $k<l$.
\item[(E2)] The anchors $\T2,\dots,\T7$ induce the same subtournament as in~$H_+$. Each original alternative relates to the anchors as $\T0$ does in~$H_+$, and each separator relates to them as $\T1$ does.
\end{enumerate}
\Cref{tab:anchors} gives an overview of the arcs created by (E2). The following observation is immediate from (E2).

\begin{table}[tbp]
\centering\small
\renewcommand{\arraystretch}{1.2}
\begin{tabular}{@{}cccc@{}}\toprule
 & \multicolumn{3}{c}{Arcs between anchor $a$ and}\\\cmidrule(l){2-4}
Anchor $a$ & other anchors & original alternative $y$ & separator $s$\\\midrule
$\T2$ & $\T2\to\T3,\T4,\T5$ & $y\to\T2$ & $\T2\to s$\\
$\T3$ & $\T3\to\T4,\T5$     & $\T3\to y$ & $s\to\T3$\\
$\T4$ & $\T4\to\T6,\T7$     & $\T4\to y$ & $\T4\to s$\\
$\T5$ & $\T5\to\T4,\T6,\T7$ & $y\to\T5$ & $s\to\T5$\\
$\T6$ & $\T6\to\T2,\T3,\T7$ & $y\to\T6$ & $\T6\to s$\\
$\T7$ & $\T7\to\T2,\T3$     & $\T7\to y$ & $s\to\T7$\\\bottomrule
\end{tabular}
\caption{The arcs involving the six anchors. Each arc between two anchors is listed once, in the row of its tail; for instance, $\T2\to\T3,\T4,\T5$ means that $\T2$ beats $\T3$, $\T4$, and~$\T5$. Original alternatives relate to the anchors as $\T0$ does in~$H_+$, and separators as $\T1$ does.}
\label{tab:anchors}
\end{table}

\begin{observation}\label{obs:copies}
Let $y$ be an original alternative and $s$ a separator. Then $y$, $s$, and $\T2,\dots,\T7$ induce a copy of $H_+$ if $y\to s$ and a copy of $H_-$ if $s\to y$, with $y$ in the role of $\T0$ and $s$ in the role of~$\T1$.
\end{observation}

\begin{lemma}\label{lem:pinning}
The tournament $T_0^{\D}$ is $3$-inducible if and only if $T_0$ is induced by a $3$-voter profile in which some voter respects~$\D$.
\end{lemma}

\begin{figure}[tbp]
\centering
\begin{tikzpicture}[
  seg/.style={draw=ink!60,line width=.4pt,minimum height=6.5mm,inner sep=2pt,font=\footnotesize,anchor=west},
  anc/.style={seg,fill=ordfill!60},
  sep/.style={seg,fill=sepfill,draw=sepdraw},
  cls/.style={seg,fill=white},
  lab/.style={font=\small,anchor=east}
]
\node[lab] at (-.15,0) {Voter $1$};
\node[anc] (a1) at (0,0) {$\T7$};
\node[cls,right=0pt of a1] (a2) {$R_1$ on $V$};
\node[anc,right=0pt of a2] (a3) {$\T6\,\T2$};
\node[sep,right=0pt of a3] (a4) {$s_1\cdots s_{K-1}$};
\node[anc,right=0pt of a4] (a5) {$\T3\,\T5\,\T4$};
\node[lab] at (-.15,-1) {Voter $2$};
\node[anc] (b1) at (0,-1) {$\T5\,\T4\,\T6$};
\node[sep,right=0pt of b1] (b2) {$s_1\cdots s_{K-1}$};
\node[anc,right=0pt of b2] (b3) {$\T7\,\T3$};
\node[cls,right=0pt of b3] (b4) {$R_2$ on $V$};
\node[anc,right=0pt of b4] (b5) {$\T2$};
\node[lab] at (-.15,-2) {Voter $3$};
\node[anc] (r1) at (0,-2) {$\T2\,\T3\,\T4$};
\node[cls,right=0pt of r1] (r2) {$D_1$};
\node[sep,right=0pt of r2] (r3) {$s_1$};
\node[cls,right=0pt of r3] (r4) {$D_2$};
\node[sep,right=0pt of r4] (r5) {$s_2$};
\node[cls,right=0pt of r5] (r6) {$\cdots$};
\node[sep,right=0pt of r6] (r7) {$s_{K-1}$};
\node[cls,right=0pt of r7] (r8) {$D_K$};
\node[anc,right=0pt of r8] (r9) {$\T5\,\T6\,\T7$};
\end{tikzpicture}
\caption{The preferences used in the proof of \Cref{lem:pinning}, from most to least preferred; anchors are gray and separators orange. They arise from $R^+$ by letting the original alternatives take the place of $\T0$ and the separators that of~$\T1$.}
\label{fig:extension}
\end{figure}

\begin{proof}
Suppose first that a $3$-voter profile induces $T_0^{\D}$. Since $T_0$ is the subtournament of $T_0^{\D}$ induced by~$V$, the profile restricted to $V$ induces~$T_0$. It remains to show that one of its voters respects~$\D$. For any original alternative~$y$, the arcs $y\to\T5\to\T4\to y$ form a directed triangle, so exactly one voter disagrees with $\T5\to\T4$ by (D2); call it voter~$i$. Now let $y\in D_k$ and $y'\in D_l$ with $k<l$. By (E1), we have $y\to s_k\to y'$. By \Cref{obs:copies}, the anchors together with $y$ and $s_k$ form a copy of~$H_+$, so voter~$i$ agrees with $y\to s_k$ by \Cref{lem:template}. Likewise, the anchors together with $y'$ and $s_k$ form a copy of~$H_-$, so voter~$i$ agrees with $s_k\to y'$. Hence, voter~$i$ prefers $y$ to~$s_k$ and $s_k$ to~$y'$, and therefore $y$ to~$y'$.

Conversely, suppose that $(R_1,R_2,R_3)$ induces $T_0$ and that, after renaming the voters if necessary, voter~$3$ respects~$\D$. We extend the preferences to the alternatives of $T_0^{\D}$ by taking the profile $R^+$ as a pattern. Writing $R_i$ also for the alternatives of $V$ listed according to~$R_i$, $R_3|_{D_k}$ for the alternatives of $D_k$ listed according to~$R_3$, and $S=s_1\,s_2\cdots s_{K-1}$, the extended preferences are
\[
\begin{array}{r@{\ \ }l}
1\colon & \T7\ \ R_1\ \ \T6\ \T2\ \ S\ \ \T3\ \T5\ \T4,\\[2pt]
2\colon & \T5\ \T4\ \T6\ \ S\ \ \T7\ \T3\ \ R_2\ \ \T2,\\[2pt]
3\colon & \T2\ \T3\ \T4\ \ R_3|_{D_1}\ s_1\ R_3|_{D_2}\ s_2\ \cdots\ s_{K-1}\ R_3|_{D_K}\ \ \T5\ \T6\ \T7,
\end{array}
\]
listed from most to least preferred (see also \Cref{fig:extension}). Since voter~$3$ respects~$\D$, the sequence $R_3|_{D_1}\cdots R_3|_{D_K}$ is just~$R_3$. We check that the extended profile induces~$T_0^{\D}$.
\begin{itemize}
\item \emph{Two original alternatives:} all three voters order $V$ as before, so these arcs are those of~$T_0$.
\item \emph{Two anchors:} the anchors appear in the same relative order as in $R^+_1$, $R^+_2$, and~$R^+_3$, so these arcs are as in~$H_+$.
\item \emph{An anchor and an original alternative or a separator:} in the preferences of each voter, every original alternative stands where $\T0$ stands in~$R^+$, and every separator where $\T1$ stands. In the preferences of voter~$3$, original alternatives and separators are interleaved, but they all lie between $\T4$ and~$\T5$, just like the adjacent pair $\T0\,\T1$ in~$R^+_3$. Hence, every original alternative relates to the anchors as $\T0$ does in~$H_+$, and every separator as $\T1$ does. This is (E2).
\item \emph{Two separators:} all three voters prefer separators with smaller indices, so $s_k\to s_l$ for $k<l$.
\item \emph{An original alternative $y$ and a separator $s_k$:} voter~$1$ prefers $y$ to~$s_k$ and voter~$2$ prefers $s_k$ to~$y$, so voter~$3$ decides. It prefers $y$ to $s_k$ exactly when $\mathit{ind}(y)\le k$, which is (E1).\qedhere
\end{itemize}
\end{proof}

\section{The Reduction} \label{sec:tournament}\label{sec:nae}

We reduce from the following variant of the satisfiability problem.

\begin{definition}[Monotone NAE-3SAT]\label{def:nae}
An instance consists of variables $x_1,\dots,x_n$ and clauses $C_1,\dots,C_m$. Each clause has the form $C_j=\NAE(x_{u_j},x_{v_j},x_{w_j})$ with $u_j<v_j<w_j$, so its three variables are distinct; none of them is negated. The question is whether one of two truth values can be assigned to each variable so that no clause receives three equal values.
\end{definition}

Monotone NAE-3SAT is NP-complete \citep{Lova73a,GaJo79a} since it is equivalent to $2$-colorability of $3$-uniform hypergraphs. The problem is invariant under exchanging the two truth values, so it does not matter which of them we call true. We denote them by $1$ and~$2$ since they will correspond to voters $1$ and~$2$ in our reduction. An \emph{assignment} is thus a function $\sigma$ from the variables to $\{1,2\}$, and it is \emph{satisfying} if no clause receives three equal values. When $j$ is clear from the context, we abbreviate $u_j,v_j,w_j$ by $u,v,w$.

For the rest of this section, we fix an instance with variables $x_1,\dots,x_n$ and clauses $C_1,\dots,C_m$, indexed in an arbitrary but fixed way. By \Cref{sec:pinning}, it suffices to construct a tournament $T$ and an ordered partition $\D$ of its alternatives such that $T$ is induced by a $3$-voter profile in which some voter respects~$\D$ if and only if the instance has a satisfying assignment; \Cref{lem:pinning} then yields the tournament~$T^{\D}$, which is $3$-inducible if and only if the instance has a satisfying assignment. We call $\D$ the \emph{reference order}, and in such a profile we always let voter~$3$ be the voter that respects it.

The reference order gives us control over voters $1$ and~$2$: by (D1), if voter~$3$ disagrees with an arc, voters $1$ and~$2$ both agree with it, and if voter~$3$ agrees with an arc, at least one of them agrees with it as well. Voters $1$ and~$2$ represent the two truth values. For every occurrence of a variable in a clause, $T$ has an \emph{occurrence alternative}, and there is a \emph{threshold}~$z$ that loses to all occurrence alternatives. Since $z$ comes last in the reference order, every occurrence alternative is preferred to~$z$ by voter~$1$ or by voter~$2$, and a variable has value $h\in\{1,2\}$ if voter~$h$ prefers all its occurrence alternatives to~$z$. To make all occurrences of a variable agree, $T$ orders them by clause while the reference order lists them in reverse. Voters $1$ and~$2$ must then both follow the clause order, so a voter that prefers the last occurrence alternative of a variable to~$z$ prefers all of them to~$z$. We call the occurrence alternatives of a variable together with $z$ its \emph{variable gadget}.

To enforce the not-all-equal condition, every clause gets four \emph{ordinary} alternatives, which lose to~$z$. As voter~$3$ ranks $z$ last, voters $1$ and~$2$ both prefer $z$ to all ordinary alternatives. So if the three variables of a clause all had value~$h$, voter~$h$ would prefer the three occurrence alternatives of the clause to its four ordinary alternatives. The arcs among the seven alternatives of a clause, its \emph{clause gadget}, rule this out as long as voter~$3$ respects a fixed order of their blocks, and they allow every other combination of values. Conversely, a satisfying assignment yields the following \emph{intended profile}.
\begin{itemize}
\item For $h\in\{1,2\}$, voter~$h$ first lists the occurrence alternatives of the variables with value~$h$, then~$z$, and then the remaining alternatives clause by clause.
\item Voter~$3$ lists the alternatives according to the reference order.
\end{itemize}
The remaining arcs, between alternatives of different clauses, are oriented as in this profile.

\subsection{Construction} \label{sec:construction}

The tournament $T$ has $7m+1$ alternatives: for each clause $C_j$ and each variable $x_i$ of~$C_j$, an occurrence alternative $a_{i,j}$ and an ordinary alternative~$b_{i,j}$; for each clause $C_j$, one more ordinary alternative~$d_j$; and the threshold~$z$. The seven alternatives $a_{i,j}$, $b_{i,j}$, and $d_j$ \emph{belong to}~$C_j$.

\begin{figure}[tbp]
\centering
\resizebox{\linewidth}{!}{\begin{tikzpicture}[
  it/.style={draw,line width=.4pt,minimum width=8.6mm,minimum height=5.2mm,inner sep=0pt,font=\footnotesize,rounded corners=1pt},
  c1/.style={it,fill=Cone!12,draw=Cone},
  c2/.style={it,fill=Ctwo!12,draw=Ctwo},
  zz/.style={it,fill=white,draw=ink,line width=.8pt},
  blk/.style={draw=ink!60,dashed,line width=.5pt,rounded corners=3pt,inner sep=2.4pt},
  grp/.style={draw=ink!55,line width=.5pt},
  lab/.style={font=\footnotesize,text=ink!85},
  arc/.style={-{Stealth[length=2.2mm]},line width=.7pt,draw=ink}
]
\node[c1] (b11) at (0.46,0) {$b_{1,1}$};
\node[c1] (b21) at (1.38,0) {$b_{2,1}$};
\node[c1] (b31) at (2.30,0) {$b_{3,1}$};
\node[c1] (a11) at (3.22,0) {$a_{1,1}$};
\node[blk,fit=(b11) (b21) (b31) (a11)] {};
\node[lab,font=\scriptsize] at (1.84,0.52) {$D_1$};
\node[c2] (b22) at (4.54,0) {$b_{2,2}$};
\node[c2] (b42) at (5.46,0) {$b_{4,2}$};
\node[c2] (b52) at (6.38,0) {$b_{5,2}$};
\node[c2] (a22) at (7.30,0) {$a_{2,2}$};
\node[blk,fit=(b22) (b42) (b52) (a22)] {};
\node[lab,font=\scriptsize] at (5.92,0.52) {$D_2$};
\node[c1] (a21) at (8.62,0) {$a_{2,1}$};
\node[blk,fit=(a21)] {};
\node[lab,font=\scriptsize] at (8.62,0.52) {$D_3$};
\node[c1] (d1) at (9.94,0) {$d_1$};
\node[c1] (a31) at (10.86,0) {$a_{3,1}$};
\node[blk,fit=(d1) (a31)] {};
\node[lab,font=\scriptsize] at (10.40,0.52) {$D_4$};
\node[c2] (a42) at (12.18,0) {$a_{4,2}$};
\node[blk,fit=(a42)] {};
\node[lab,font=\scriptsize] at (12.18,0.52) {$D_5$};
\node[c2] (d2) at (13.50,0) {$d_2$};
\node[c2] (a52) at (14.42,0) {$a_{5,2}$};
\node[blk,fit=(d2) (a52)] {};
\node[lab,font=\scriptsize] at (13.96,0.52) {$D_6$};
\node[zz] (z) at (15.74,0) {$z$};
\node[blk,fit=(z)] {};
\node[lab,font=\scriptsize] at (15.74,0.52) {$D_7$};
\draw[grp] (0.04,0.78) -- (0.04,0.86) -- (3.64,0.86) -- (3.64,0.78);
\node[lab,anchor=south] at (1.84,0.88) {$x_1$};
\draw[grp] (4.12,0.78) -- (4.12,0.86) -- (9.04,0.86) -- (9.04,0.78);
\node[lab,anchor=south] at (6.58,0.88) {$x_2$};
\draw[grp] (9.52,0.78) -- (9.52,0.86) -- (11.28,0.86) -- (11.28,0.78);
\node[lab,anchor=south] at (10.40,0.88) {$x_3$};
\draw[grp] (11.76,0.78) -- (11.76,0.86) -- (12.60,0.86) -- (12.60,0.78);
\node[lab,anchor=south] at (12.18,0.88) {$x_4$};
\draw[grp] (13.08,0.78) -- (13.08,0.86) -- (14.84,0.86) -- (14.84,0.78);
\node[lab,anchor=south] at (13.96,0.88) {$x_5$};
\node[lab,anchor=south,font=\scriptsize] at (6.58,1.22) {first in $C_2$, then in $C_1$};
\draw[arc] (a21.south) .. controls +(0,-0.95) and +(0,-0.95) .. (a22.south);
\node[lab,anchor=east] at (7.18,-0.80) {(V1): $a_{2,1}\to a_{2,2}$};
\node[c1,minimum width=3mm,minimum height=3mm] (l1) at (11.72,-0.75) {};
\node[lab,anchor=west] at (l1.east) {alternative of $C_1$};
\node[c2,minimum width=3mm,minimum height=3mm] (l2) at (14.82,-0.75) {};
\node[lab,anchor=west] at (l2.east) {alternative of $C_2$};
\end{tikzpicture}}
\caption{The reference order $\D=(D_1,\dots,D_7)$ for the running example $C_1=\NAE(x_1,x_2,x_3)$ and $C_2=\NAE(x_2,x_4,x_5)$, from left to right. Each dashed box is a block; every block other than~$\{z\}$ belongs to one occurrence of a variable in a clause. The blocks are grouped by variable, and the two blocks of~$x_2$ appear in decreasing clause order, so the arc $a_{2,1}\to a_{2,2}$ from (V1) points backward (R2). Within each clause, the blocks follow the order of its variables (R1), and $z$ comes last (R3).}
\label{fig:blocks}
\end{figure}
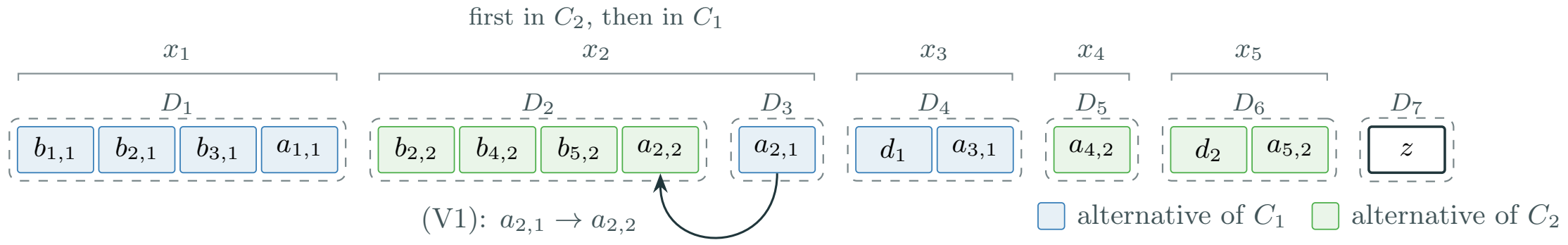

The reference order has one block for each occurrence of a variable in a clause and the block~$\{z\}$. For a clause $C_j=\NAE(x_u,x_v,x_w)$ with $u<v<w$, the blocks of the occurrences of $x_u$, $x_v$, and $x_w$ are
\[
B_{u,j}=\{b_{u,j},\,b_{v,j},\,b_{w,j},\,a_{u,j}\},\qquad
B_{v,j}=\{a_{v,j}\},\qquad
B_{w,j}=\{d_j,\,a_{w,j}\}.
\]
The blocks are sorted by variable, the blocks of the same variable by decreasing clause index, and $\{z\}$ comes last; this gives $\D=(D_1,\dots,D_{3m+1})$. \Cref{fig:blocks} shows $\D$ for the running example $C_1=\NAE(x_1,x_2,x_3)$ and $C_2=\NAE(x_2,x_4,x_5)$. We will use three properties of~$\D$.
\begin{enumerate}[leftmargin=3.2em]
\item[(R1)] The blocks of a clause $C_j$ appear in the order $B_{u,j},B_{v,j},B_{w,j}$ because $u<v<w$.
\item[(R2)] The occurrence alternatives of a variable appear in decreasing clause order.
\item[(R3)] The threshold $z$ comes last.
\end{enumerate}

\begin{figure}[tbp]
\centering
\resizebox{\linewidth}{!}{\begin{tikzpicture}[
  cand/.style={circle,draw=ink,line width=.5pt,minimum size=8.4mm,inner sep=0pt,font=\small},
  ord/.style={cand,fill=ordfill},
  mk/.style={cand,fill=white,draw=accent,line width=.8pt},
  arc/.style={-Latex,line width=.6pt,draw=ink!75},
  oarc/.style={arc,draw=accent,line width=.8pt},
  cell/.style={draw=ink!60,line width=.4pt,minimum width=7mm,minimum height=6mm,inner sep=0pt,font=\small},
  co/.style={cell,fill=ordfill},
  cA/.style={cell,fill=Afill,draw=Adraw},
  cB/.style={cell,fill=Bfill,draw=Bdraw},
  sepline/.style={line width=1.2pt,draw=ink},
  lab/.style={font=\footnotesize,text=ink!80}
]
\node[font=\small,anchor=west] at (-0.7,3.75) {\textbf{(a)} The gadget of $\NAE(x_u,x_v,x_w)$};
\node[ord] (bu) at (0,0) {$b_u$};
\node[ord] (bw) at (1.8,0) {$b_w$};
\node[ord] (bv) at (3.6,0) {$b_v$};
\node[ord] (dj) at (1.8,-1.8) {$d_j$};
\node[mk] (au) at (0.9,2) {$a_u$};
\node[mk] (aw) at (2.7,2) {$a_w$};
\node[mk] (av) at (4.5,2) {$a_v$};
\node[lab] at (0.9,3.1) {$x_u$};
\node[lab] at (2.7,3.1) {$x_w$};
\node[lab] at (4.5,3.1) {$x_v$};
\draw[arc] (bw) -- (bu);
\draw[arc] (bw) -- (bv);
\draw[arc] (bu) -- (dj);
\draw[arc] (bv) -- (dj);
\draw[arc] (dj) -- (bw);
\draw[arc] (bv) .. controls (4.6,-3.1) and (-1.0,-3.1) .. (bu);
\draw[arc] (au) -- (aw);
\draw[arc] (av) -- (aw);
\draw[arc] (au) to[bend left=32] (av);
\draw[oarc] (bu) -- (au);
\draw[oarc] (bw) -- (au);
\draw[oarc] (bw) -- (aw);
\draw[oarc] (bv) -- (aw);
\draw[oarc] (bv) -- (av);
\begin{scope}[xshift=7.1cm]
\node[font=\small,anchor=west] at (-0.6,3.75) {\textbf{(b)} A clause profile when only $x_u$ has value $2$};
\def\w{0.74}
\foreach \y/\name/\row in {2.2/1/{a_v/cA,a_w/cA,b_u/co,d_j/co,b_w/co,a_u/cB,b_v/co},
                           1.2/2/{a_u/cB,b_v/co,d_j/co,a_v/cA,b_w/co,a_w/cA,b_u/co},
                           0.2/3/{b_w/co,b_v/co,b_u/co,a_u/cB,a_v/cA,a_w/cA,d_j/co}}{
  \node[font=\small,anchor=east] at (0.3,\y) {Voter $\name$};
  \foreach \l/\st [count=\i] in \row { \node[\st] at (\i*\w+0.3,\y) {$\l$}; }
}
\draw[sepline] (2.5*\w+0.3,1.85) -- (2.5*\w+0.3,2.55);
\draw[sepline] (1.5*\w+0.3,0.85) -- (1.5*\w+0.3,1.55);
\draw[sepline] (4.5*\w+0.3,-0.15) -- (4.5*\w+0.3,0.55);
\draw[sepline] (5.5*\w+0.3,-0.15) -- (5.5*\w+0.3,0.55);
\node[cA,minimum width=3mm,minimum height=3mm] (lA) at (-0.4,-0.9) {};
\node[lab,anchor=west] at (lA.east) {value $1$};
\node[cB,minimum width=3mm,minimum height=3mm] (lB) at (1.45,-0.9) {};
\node[lab,anchor=west] at (lB.east) {value $2$};
\node[co,minimum width=3mm,minimum height=3mm] (lO) at (3.3,-0.9) {};
\node[lab,anchor=west] at (lO.east) {ordinary};
\draw[sepline] (-0.4,-1.6) -- (-0.4,-1.3);
\node[lab,anchor=west] at (-0.3,-1.45) {end of prefix / block};
\end{scope}
\end{tikzpicture}}
\caption{The clause gadget of $C_j=\NAE(x_u,x_v,x_w)$, with the second index dropped.
(a)~The arcs~(C). Green arcs point from an ordinary to an occurrence alternative; in all
other pairs of an ordinary and an occurrence alternative, the occurrence alternative wins.
(b)~A clause profile that induces (a) when only $x_u$ has value~$2$. Voters $1$ and~$2$ start
with the occurrence alternatives of their value, and voter~$3$ respects the three blocks
of~$C_j$; the bars mark these parts.}
\label{fig:gadget}
\end{figure}
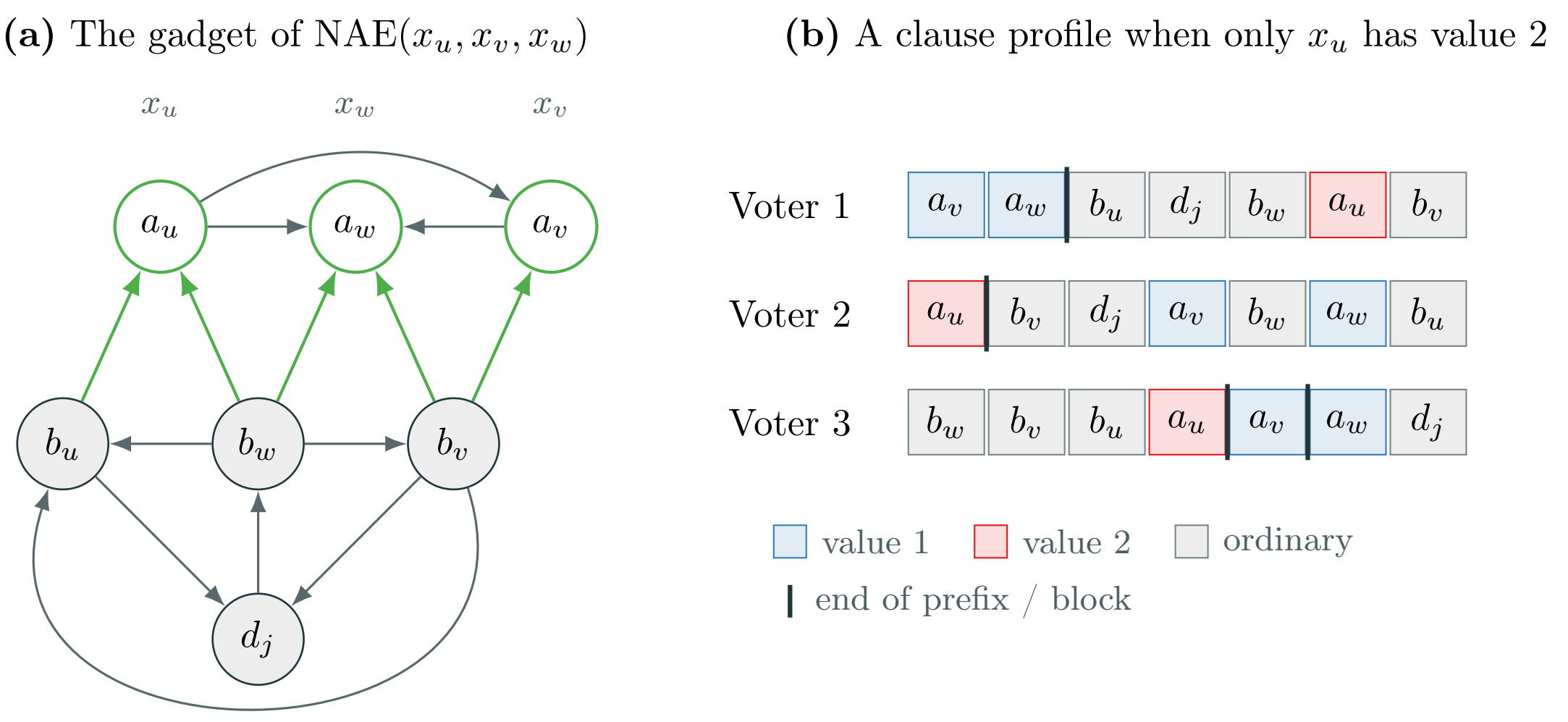

The arcs of $T$ are given by the following rules, which \Cref{tab:arcs} summarizes.
\begin{enumerate}[leftmargin=3.2em]
\item[(C)] Inside a clause $C_j=\NAE(x_u,x_v,x_w)$, where we write $a_u,\dots,b_w$ for $a_{u,j},\dots,b_{w,j}$ (\Cref{fig:gadget}(a)): the occurrence alternatives are ordered by variable, $a_u\to a_v\to a_w$ and $a_u\to a_w$; among the ordinary alternatives, $b_w\to b_v\to b_u$ and $b_w\to b_u$, and $d_j$ beats $b_w$ but loses to $b_u$ and~$b_v$; and an ordinary alternative beats an occurrence alternative only in the five cases $b_u\to a_u$, $b_w\to a_u$, $b_v\to a_v$, $b_v\to a_w$, and $b_w\to a_w$.
\item[(V1)] The occurrence alternatives of each variable are ordered by clause: $a_{i,j}\to a_{i,j'}$ whenever $j<j'$.
\item[(V2)] Every occurrence alternative beats~$z$, and $z$ beats every ordinary alternative.
\item[(O)] Let $y$ belong to $C_j$ and $y'$ to $C_{j'}$ with $j<j'$, where $y$ and $y'$ are not occurrence alternatives of the same variable. If $y'$ is ordinary, then $y\to y'$. Otherwise, the one of $y$ and $y'$ in the earlier block wins.
\end{enumerate}
Rule (C) defines the clause gadgets, rules (V1) and (V2) make the variable gadgets work, and rule (O) orients the remaining pairs as the intended profile does. By (R2), the arcs (V1) run against the reference order. Since the blocks are sorted by variable, (C) and (O) imply that for $i<k$, every occurrence alternative of $x_i$ beats every occurrence alternative of~$x_k$.

\begin{table}[tbp]
\centering\small
\begin{tabularx}{\linewidth}{@{}>{\raggedright\arraybackslash}p{.22\linewidth}>{\raggedright\arraybackslash}p{.34\linewidth}>{\raggedright\arraybackslash}X c@{}}\toprule
Alternative $y$ & Alternative $y'$ & Winner & Rule\\\midrule
in $C_j$ & in the same clause $C_j$ & as in the clause gadget (\Cref{fig:gadget}) & (C)\\\midrule
occurrence $a_{i,j}$ & occurrence $a_{i,j'}$, $j<j'$ & $y$ & (V1)\\
in $C_j$ & ordinary, in $C_{j'}$, $j<j'$ & $y$ & (O)\\
in $C_j$ & occurrence, in $C_{j'}$, $j<j'$, if not covered by (V1) & the one in the earlier block & (O)\\\midrule
occurrence & $z$ & $y$ & (V2)\\
ordinary & $z$ & $z$ & (V2)\\\bottomrule
\end{tabularx}
\caption{Overview of the arcs of~$T$.}
\label{tab:arcs}
\end{table}

\subsection{Correctness} \label{sec:correctness}

We first analyze a single clause gadget, then show that the variable gadgets assign consistent values, and finally prove both directions of the reduction.

Fix a clause $C_j=\NAE(x_u,x_v,x_w)$ and write $a_u$, $a_v$, $a_w$, $b_u$, $b_v$, $b_w$ for its occurrence and ordinary alternatives, as in~(C). We call a profile on the seven alternatives of $C_j$ a \emph{clause profile} if it induces the arcs (C) and voter~$3$ respects the blocks $\bigl(B_{u,j},B_{v,j},B_{w,j}\bigr)$. By (R1), every profile that induces $T$ and in which voter~$3$ respects~$\D$ yields a clause profile when restricted to the alternatives of~$C_j$.

\begin{lemma}\label{lem:gadget}
\begin{enumerate}[label=(\alph*)]
\item In no clause profile does voter~$1$ or voter~$2$ prefer all three occurrence alternatives to all four ordinary alternatives.
\item Let each of $a_u,a_v,a_w$ receive a value in $\{1,2\}$ such that not all three values are equal. Then there is a clause profile in which, for $h\in\{1,2\}$, the preferences of voter~$h$ start with the occurrence alternatives of value~$h$, in the order $a_u,a_v,a_w$. We call this the \emph{prefix property}.
\end{enumerate}
\end{lemma}

\begin{proof}
(a) By the symmetry between voters $1$ and~$2$, it suffices to consider voter~$2$. Suppose for the sake of contradiction that voter~$2$ prefers $a_u,a_v,a_w$ to all ordinary alternatives. We show that the preferences of voter~$2$ would then be cyclic.

Voter~$2$ disagrees with the five arcs from an ordinary to an occurrence alternative, so by (D1), voters $1$ and~$3$ agree with them. Since voter~$3$ respects the blocks of~$C_j$, it prefers $b_u,b_v,b_w$ to $a_v$ and~$a_w$, so it disagrees with $a_v\to b_u$, $a_v\to b_w$, and $a_w\to b_u$, and by (D1), voter~$1$ agrees with them.

From these eight arcs, we obtain
\begin{equation}\label{eq:voter1}
b_v\Ri{1}a_v\Ri{1}b_w\Ri{1}a_w\Ri{1}b_u\qquad\text{and}\qquad b_w\Ri{1}a_u.
\end{equation}
Next, we consider voter~$3$. By~\eqref{eq:voter1}, voter~$1$ disagrees with $a_u\to b_v$, so voter~$3$ prefers $a_u$ to~$b_v$; since voter~$3$ agrees with $b_u\to a_u$, we have $b_u\Ri{3}a_u\Ri{3}b_v$. Since voter~$3$ respects the blocks of~$C_j$, we also have $b_w\Ri{3}d_j$. Hence, voter~$3$ disagrees with $b_v\to b_u$ and with $d_j\to b_w$, and voters $1$ and~$2$ both prefer $b_v$ to $b_u$ and $d_j$ to~$b_w$.

Finally, \eqref{eq:voter1} gives $b_v\Ri{1}b_w$, and together with $d_j\Ri{1}b_w$, it gives $d_j\Ri{1}b_w\Ri{1}b_u$, so voter~$1$ disagrees with $b_w\to b_v$ and with $b_u\to d_j$, and voter~$2$ agrees with both. Thus, $d_j\Ri{2}b_w\Ri{2}b_v\Ri{2}b_u\Ri{2}d_j$, which is impossible.

(b) Exchanging the roles of voters $1$ and~$2$ if necessary, exactly one occurrence alternative has value~$2$. \Cref{tab:gadget} lists suitable profiles for each of the three possibilities, and \Cref{fig:gadget}(b) shows the first one. That they induce (C), that voter~$3$ respects the blocks, and the prefix property can be checked directly.\footnote{We also verified these profiles by computer.}
\end{proof}

\begin{table}[tbp]
\centering\small
\begin{tabular}{@{}cccc@{}}\toprule
Occurrence with value $2$ & $R_1$ & $R_2$ & $R_3$\\\midrule
$a_u$ & $a_v\,a_w\,b_u\,d_j\,b_w\,a_u\,b_v$ & $a_u\,b_v\,d_j\,a_v\,b_w\,a_w\,b_u$ & $b_w\,b_v\,b_u\,a_u\,a_v\,a_w\,d_j$\\
$a_v$ & $a_u\,a_w\,b_v\,a_v\,b_u\,d_j\,b_w$ & $a_v\,d_j\,b_w\,b_v\,a_w\,b_u\,a_u$ & $b_w\,b_u\,a_u\,b_v\,a_v\,a_w\,d_j$\\
$a_w$ & $a_u\,a_v\,d_j\,b_w\,b_v\,a_w\,b_u$ & $a_w\,b_v\,a_v\,b_u\,d_j\,b_w\,a_u$ & $b_w\,b_u\,a_u\,b_v\,a_v\,a_w\,d_j$\\\bottomrule
\end{tabular}
\caption{Clause profiles for \Cref{lem:gadget}(b) when exactly one occurrence alternative (first column) has value~$2$. Preferences are listed from most to least preferred.}
\label{tab:gadget}
\end{table}

Next, we show that every variable receives a value, and that this value is the same in all its clauses.

\begin{lemma}\label{lem:variable}
Let a $3$-voter profile induce~$T$, and suppose that voter~$3$ respects~$\D$.
\begin{enumerate}[label=(\alph*)]
\item Voters $1$ and~$2$ prefer $z$ to every ordinary alternative.
\item For every variable $x_i$ that occurs in some clause, voter~$1$ or voter~$2$ prefers all occurrence alternatives of $x_i$ to~$z$, and hence to every ordinary alternative.
\end{enumerate}
\end{lemma}

\begin{proof}
(a) By (R3), $z$ is the least preferred alternative of voter~$3$. Since $z$ beats every ordinary alternative by (V2), voter~$3$ disagrees with all these arcs, so by (D1), voters $1$ and~$2$ both prefer $z$ to every ordinary alternative.

(b) Let $g_1,\dots,g_t$ be the occurrence alternatives of $x_i$ in increasing clause order. By (V1), $g_k\to g_l$ for $k<l$, whereas voter~$3$ prefers them in decreasing clause order by (R2). So voter~$3$ disagrees with all these arcs, and by (D1), voters $1$ and~$2$ both prefer $g_k$ to $g_l$ whenever $k<l$. The arc $g_t\to z$ from (V2) needs two supporters, and voter~$3$ is one of them, so at least one of voters $1$ and~$2$, say voter~$h$, prefers $g_t$ to~$z$. Together with~(a), this gives
\[
g_1\Ri{h}\cdots\Ri{h}g_t\Ri{h}z\Ri{h}o\qquad\text{for every ordinary alternative }o.\qedhere
\]
\end{proof}

We can now prove both directions of the reduction.

\begin{lemma}\label{lem:complete}
If the instance has a satisfying assignment, then $T$ is induced by a $3$-voter profile in which voter~$3$ respects~$\D$.
\end{lemma}

\begin{proof}
Fix a satisfying assignment~$\sigma$. By \Cref{lem:gadget}(b), there is, for every clause~$C_j$, a clause profile $R^j$ that has the prefix property for the values given by~$\sigma$. We write $R^j_1$, $R^j_2$, and $R^j_3$ for the preferences of the three voters in~$R^j$. We combine them into the intended profile described at the beginning of this section.
\begin{itemize}
\item The preferences $R_3$ list the blocks $D_1,\dots,D_{3m+1}$ in this order and order each block other than $\{z\}$ as $R^j_3$, where $C_j$ is the clause to which the block belongs. In particular, voter~$3$ respects~$\D$.
\item For $h\in\{1,2\}$, the preferences $R_h$ begin with the occurrence alternatives of all variables of value~$h$, ordered by variable and then by clause; we call this initial segment the \emph{head} of~$R_h$. Then comes~$z$, followed by $R^1_h,\dots,R^m_h$ in this order, each with the alternatives of the head removed.
\end{itemize}
The example after this proof shows this profile for the running example. We check that $R=(R_1,R_2,R_3)$ induces~$T$.

\emph{Pairs within a clause.} Let $h\in\{1,2\}$. By the prefix property, the occurrence alternatives removed from $R^j_h$ form an initial segment of it, in the order $a_{u,j},a_{v,j},a_{w,j}$, which is also their order in the head. Hence, $R_h$ restricted to $C_j$ is~$R^j_h$. By (R1), $R_3$ restricted to $C_j$ is~$R^j_3$. Since $R^j$ induces the arcs (C), the arcs inside $C_j$ are correct.

\emph{Pairs involving $z$.} Voters $1$ and~$2$ both prefer $z$ to every ordinary alternative. An occurrence alternative of a variable of value~$h$ lies in the head of~$R_h$ and is preferred to $z$ by voter~$3$, so it beats~$z$. This is (V2).

\emph{Pairs from different clauses.} Let $y$ belong to $C_j$ and $y'$ to $C_{j'}$ with $j<j'$. Outside its head, each $R_h$ with $h\in\{1,2\}$ lists the clauses in increasing order, so voter~$h$ prefers $y'$ to $y$ only if $y'$ lies in the head of~$R_h$ and $y$ does not precede $y'$ there. If $y'$ is ordinary, it lies in no head, so voters $1$ and~$2$ both prefer $y$ to~$y'$, as (O) requires. Otherwise, $y'$ lies in the head of exactly one~$R_h$, and the other voter in $\{1,2\}$ prefers $y$ to~$y'$. Voter~$h$ does so as well if $y$ is an occurrence alternative of the same variable as~$y'$, or of a smaller variable of value~$h$; then $y\to y'$ by (V1) or by (O), respectively, since in the second case $y$ lies in an earlier block. In all remaining cases, $y$ and $y'$ are not occurrence alternatives of the same variable and voters $1$ and~$2$ disagree, so voter~$3$ decides according to the reference order, as (O) requires.
\end{proof}

\begin{example}
For the running example of \Cref{fig:blocks} and the assignment with $\sigma(x_1)=\sigma(x_3)=\sigma(x_4)=1$ and $\sigma(x_2)=\sigma(x_5)=2$, the proof of \Cref{lem:complete} yields the following profile; bars mark the heads and, for voter~$3$, the blocks.
\[\small
\begin{array}{r@{\ \ }l}
R_1\colon & a_{1,1}\,a_{3,1}\,a_{4,2}\mid z\mid b_{2,1}\,a_{2,1}\,b_{1,1}\,d_1\,b_{3,1}\;\;d_2\,b_{5,2}\,b_{4,2}\,a_{5,2}\,b_{2,2}\,a_{2,2}\\
R_2\colon & a_{2,1}\,a_{2,2}\,a_{5,2}\mid z\mid d_1\,b_{3,1}\,b_{2,1}\,a_{3,1}\,b_{1,1}\,a_{1,1}\;\;b_{4,2}\,a_{4,2}\,b_{2,2}\,d_2\,b_{5,2}\\
R_3\colon & b_{3,1}\,b_{1,1}\,a_{1,1}\,b_{2,1}\mid b_{5,2}\,b_{2,2}\,a_{2,2}\,b_{4,2}\mid a_{2,1}\mid a_{3,1}\,d_1\mid a_{4,2}\mid a_{5,2}\,d_2\mid z
\end{array}
\]
This profile induces~$T$. For instance, both occurrences of $x_2$ are in the head of $R_2$ in clause order, while voter~$3$ prefers $a_{2,2}$ to~$a_{2,1}$.
\end{example}

\begin{lemma}\label{lem:sound}
If $T$ is induced by a $3$-voter profile in which some voter respects~$\D$, then the instance has a satisfying assignment.
\end{lemma}

\begin{proof}
Consider such a profile and assume without loss of generality that voter~$3$ respects~$\D$. We set $\sigma(x_i)=1$ if voter~$1$ prefers all occurrence alternatives of $x_i$ to~$z$, and $\sigma(x_i)=2$ otherwise. By \Cref{lem:variable}, if $x_i$ occurs in some clause, voter~$\sigma(x_i)$ prefers all occurrence alternatives of $x_i$ to every ordinary alternative.

Suppose for the sake of contradiction that some clause $C_j=\NAE(x_u,x_v,x_w)$ is not satisfied, that is, $\sigma(x_u)=\sigma(x_v)=\sigma(x_w)=h$ for some $h\in\{1,2\}$. Then voter~$h$ prefers all three occurrence alternatives of $C_j$ to all four ordinary alternatives of~$C_j$. Since the profile restricted to the alternatives of $C_j$ is a clause profile, this contradicts \Cref{lem:gadget}(a). Hence, $\sigma$ is a satisfying assignment.
\end{proof}

Together, the lemmas yield our main result, which we restate for convenience.

\mainthm*

\begin{proof}
By \Cref{lem:complete,lem:sound}, $T$ is induced by a $3$-voter profile in which some voter respects $\D$ if and only if the instance has a satisfying assignment. By \Cref{lem:pinning}, this holds if and only if $T^{\D}$ is $3$-inducible. The tournament $T^{\D}$ has $(7m+1)+3m+6=10m+7$ alternatives and can be constructed in polynomial time. Together with membership in NP (\Cref{sec:majority}), this proves the theorem.
\end{proof}

\section*{Acknowledgments}

This material is based on work supported by the Deutsche Forschungsgemeinschaft under grant {BR~2312/14-1}. The reduction proof for $3$-inducibility was found by ChatGPT Astra 6.2, the extensions to $k>3$ by Claude Fable 5.2. We simplified the construction and improved the exposition, with some support from Claude Opus 5.5.

\appendix

\section{More Than Three Voters} \label{app:more-voters}

In this appendix, we prove \Cref{thm:more-voters}, which extends \Cref{thm:main} to every fixed number of voters. The gadgets of the main body rely on (D1)--(D3), which have no analogue for more than three voters. Instead, we pad a $3$-voter profile with voters whose preferences are fixed up to reversal, and we enforce this shape by large grids of auxiliary alternatives, whose size is determined by Ramsey's theorem \citep{Rams30a}.

For an even number of voters, majority digraphs need not be complete. Let $R=(R_1,\dots,R_k)$ be a profile of $k$ voters. The \emph{margin} of two alternatives $y$ and $y'$ is $\mu_R(y,y')=|\{i\colon y\Ri{i}y'\}|-|\{i\colon y'\Ri{i}y\}|$. The profile $R$ \emph{induces} the digraph $G=(V,E)$ with $(y,y')\in E$ if and only if $\mu_R(y,y')>0$, and we again write $y\to y'$ for such an arc. If $\mu_R(y,y')=0$, the pair is \emph{tied}. Since $\mu_R(y,y')\equiv k\pmod 2$, ties occur only for even~$k$, and the digraph induced by an odd number of voters is a tournament. A digraph is \emph{$k$-inducible} if it is induced by some $k$-voter profile. Deciding $k$-inducibility is in NP since a profile serves as a certificate, and for $k\le2$ it can be done in polynomial time (see \Cref{sec:intro}).

\morevotersthm*

\subsection{Grid Digraphs} \label{app:grids}

The following simple fact takes the place of (D1)--(D3).

\begin{lemma}\label{lem:switch}
Let $R$ be a profile, and let $x,y,x',y'$ be alternatives with $\mu_R(x,y)>\mu_R(x',y')$. Then the number of voters that prefer $x$ to $y$ and $y'$ to $x'$ exceeds the number of voters that prefer $y$ to $x$ and $x'$ to $y'$. In particular, if $x\to y$ is an arc of the induced digraph and $x'\to y'$ is not, then some voter prefers $x$ to $y$ and $y'$ to~$x'$.
\end{lemma}

\begin{proof}
Let $n_{++}$, $n_{+-}$, $n_{-+}$, and $n_{--}$ count the voters by whether they prefer $x$ to $y$ (first sign) and $x'$ to $y'$ (second sign). Then $\mu_R(x,y)-\mu_R(x',y')=(n_{++}+n_{+-}-n_{-+}-n_{--})-(n_{++}+n_{-+}-n_{+-}-n_{--})=2(n_{+-}-n_{-+})$.
\end{proof}

Fix $d\ge3$. For points $x,y\in\mathbb Q^d$ that differ in every coordinate, the \emph{sign vector} $\varepsilon(x,y)\in\{+,-\}^d$ has $\varepsilon_i(x,y)=+$ if and only if $x_i<y_i$. Let $P\subseteq\mathbb Q^d$ be a finite set of \emph{marked} points; different marked points may share coordinates. On axis~$i$, the $i$-th coordinates of the marked points divide the remaining rational numbers into open intervals, the \emph{cells} of axis~$i$, including the two unbounded ones. A \emph{box} $B=(B_1,\dots,B_d)$ consists of one cell of every axis. A \emph{grid with $L$ levels per cell} chooses $L$ rational numbers, the \emph{levels}, in every cell; its \emph{auxiliary points} are all points whose $i$-th coordinate is a level on axis~$i$, for every~$i$. We write $\Gamma$ for the set of auxiliary points and $\Gamma_B$ for those in box~$B$. Every auxiliary point differs from every marked point in every coordinate.

Fix a numbering of~$\Gamma$, and let $\lhd_i$ order points from the smallest to the largest $i$-th coordinate, where auxiliary points with the same $i$-th coordinate are ordered by the numbering. Then $\lhd_i$ orders every pair that contains an auxiliary point. A digraph on $P\cup\Gamma$ is a \emph{grid digraph} if every pair that contains an auxiliary point is oriented or tied as in the digraph induced by $(\lhd_1,\dots,\lhd_d)$. Pairs of marked points are arbitrary. In particular, if $x$ is an auxiliary point and $y$ a point that differs from $x$ in every coordinate, then $x\to y$ if $\varepsilon(x,y)$ has more than $d/2$ plus signs, and $x\to y$ does not hold otherwise. The intended profile lets voter~$i$ rank the alternatives according to~$\lhd_i$. The next lemma shows that, as far as marked points are concerned, a sufficiently large grid admits no other profile.

\begin{lemma}\label{lem:grid}
For all $d\ge3$ and~$\ell$, there is an $L=L(d,\ell)$ such that the following holds. Let $|P|\le\ell$, and let a $d$-voter profile induce a grid digraph with $L$ levels per cell. Then the voters can be numbered $v_1,\dots,v_d$ such that, for all $i$ and all $p,q\in P$ with $p_i<q_i$, voter~$v_i$ prefers $p$ to~$q$.
\end{lemma}

\begin{proof}
\emph{Step 0: a regular subgrid.} Let $J_1,\dots,J_M$ be the cells of all axes, so $M\le d(\ell+1)$, and let $\lambda_J(1)<\dots<\lambda_J(L)$ be the levels of cell~$J$. For a box $B$ and $\alpha\in\{1,\dots,L\}^d$, let $B[\alpha]$ be the auxiliary point with coordinates $\lambda_{B_i}(\alpha_i)$. We claim that, if $L$ is large enough in terms of $d$ and~$\ell$, we can keep five levels in every cell (Step~3 needs a middle level with two levels on each side) such that the remaining subgrid satisfies the following two conditions for every voter~$v$.
\begin{enumerate}[label=\textit{(G\arabic*)},leftmargin=3.2em]
\item For boxes $B$ and $C$ and points $x\in\Gamma_B$ and $y\in\Gamma_C$ that differ in every coordinate, whether $x\Ri{v}y$ depends only on $B$, $C$, and $\varepsilon(x,y)$.
\item For $p\in P$ and a box~$B$, voter~$v$ prefers either all of $\Gamma_B$ to $p$ or $p$ to all of~$\Gamma_B$.
\end{enumerate}
To see this, consider the following colorings of subsets of $\{1,\dots,L\}$. For boxes $B$ and $C$ and a bijection $\tau$ from $\{1,2\}\times\{1,\dots,d\}$ to $\{1,\dots,2d\}$, color a set $s_1<\dots<s_{2d}$ by the list, over all voters~$v$, of whether $v$ prefers $B[(s_{\tau(1,i)})_i]$ to $C[(s_{\tau(2,i)})_i]$. For $p\in P$, a box $B$, and a bijection $\tau$ of $\{1,\dots,d\}$, color a set $s_1<\dots<s_d$ by the list of whether $v$ prefers $B[(s_{\tau(i)})_i]$ to~$p$. These are finitely many colorings with $2^d$ colors each, and their number depends only on $d$ and~$\ell$. Applying Ramsey's theorem once for each coloring yields, for large~$L$, a set $I$ of $5M$ indices on which every one of these colorings is constant. We split $I$ into consecutive groups $I_1<\dots<I_M$ of five indices and keep in cell $J_t$ the levels with index in~$I_t$.

For (G1), let $x=B[\alpha]$ and $y=C[\beta]$ be points of the subgrid that differ in every coordinate. The $2d$ indices $\alpha_1,\dots,\alpha_d,\beta_1,\dots,\beta_d$ are distinct: indices of different cells lie in different groups, and $\alpha_i$ and $\beta_i$ belong to the same cell only if $B_i=C_i$, in which case $\alpha_i\ne\beta_i$ because $x_i\ne y_i$. Their relative order is determined by $B$, $C$, and $\varepsilon(x,y)$: across different cells by the numbering of the cells, and within a cell by $\varepsilon_i(x,y)$. So the color of $\{\alpha_1,\dots,\beta_d\}$ under the coloring for $B$, $C$, and the resulting~$\tau$ records whether each voter prefers $x$ to~$y$, and (G1) follows. The argument for (G2) is the same with $d$ indices. From now on, $\Gamma$ denotes the subgrid; restricted to it, the profile still induces a grid digraph.

\emph{Step 1: in each box, every voter follows one coordinate, up to sign.} Fix a voter $v$ and a box~$B$. By (G1), there is $f\colon\{+,-\}^d\to\{+,-\}$ such that $x\Ri{v}y$ if and only if $f(\varepsilon(x,y))=+$, for all $x,y\in\Gamma_B$ that differ in every coordinate. Clearly, $f(-\varepsilon)=-f(\varepsilon)$. Call $S\subseteq\{1,\dots,d\}$ \emph{winning} if $f(\varepsilon^S)=f(+,\dots,+)$, where $\varepsilon^S$ is $+$ exactly on~$S$. Exactly one of $S$ and its complement is winning, so $\emptyset$ is not. If $S$ and $S'$ are winning, so is $S\cap S'$: choose $x,y,z\in\Gamma_B$ whose levels are ordered $x<y<z$ on the axes in $S\cap S'$, $z<x<y$ on $S\setminus S'$, $y<z<x$ on $S'\setminus S$, and $z<y<x$ on all other axes. Then $\varepsilon(x,y)=\varepsilon^S$, $\varepsilon(y,z)=\varepsilon^{S'}$, and $\varepsilon(x,z)=\varepsilon^{S\cap S'}$, and the claim follows from transitivity. If no singleton were winning, all complements of singletons would be winning, and so would be their intersection~$\emptyset$. So some $\{c\}$ is winning; since every winning set meets $\{c\}$, a set is winning if and only if it contains~$c$. Hence, $f(\varepsilon)=\varepsilon_c$ or $f(\varepsilon)=-\varepsilon_c$, depending on $f(+,\dots,+)$, and we say that $v$ follows coordinate $c$ in $B$ with positive or negative sign, respectively.

\emph{Step 2: in each box, the voters follow different coordinates, all with positive sign.} Fix a box $B$ and a coordinate~$i$, and let $g^+_i$ and $g^-_i$ be the numbers of voters that follow coordinate $i$ in $B$ with positive and with negative sign. Choose $\varepsilon\in\{+,-\}^d$ with $\varepsilon_i=+$ and exactly $\lfloor d/2\rfloor$ further plus signs, and let $\varepsilon'$ arise from $\varepsilon$ by changing~$\varepsilon_i$. For $x,y,x',y'\in\Gamma_B$ with $\varepsilon(x,y)=\varepsilon$ and $\varepsilon(x',y')=\varepsilon'$, the grid digraph contains $x\to y$ but not $x'\to y'$. A voter that follows a coordinate other than $i$ treats both pairs alike. So the voters that prefer $x$ to $y$ and $y'$ to $x'$ are those counted by~$g^+_i$, and the voters that prefer $y$ to $x$ and $x'$ to $y'$ are those counted by~$g^-_i$. By \Cref{lem:switch}, $g^+_i-g^-_i\ge1$. Summing over $i$ gives $d=\sum_i(g_i^++g_i^-)\ge d+2\sum_ig_i^-$, so $g_i^-=0$ and $g_i^+=1$ for all~$i$. Thus, for every box~$B$, there is a bijection $v\mapsto c_v(B)$ from the voters to the coordinates such that each voter~$v$ follows coordinate $c_v(B)$ in $B$ with positive sign.

\emph{Step 3: neighboring boxes.} Let $B$ and $C$ be boxes that differ in exactly one axis~$j$, with $B_j$ below~$C_j$. For $x\in\Gamma_B$ and $y\in\Gamma_C$ that differ in every coordinate, we have $\varepsilon_j(x,y)=+$, so by (G1) there is a function $g_v$ such that $x\Ri{v}y$ if and only if $g_v(\rho)=+$, where the \emph{pattern} $\rho$ of $x$ against~$y$ is $\varepsilon(x,y)$ without its $j$-th entry.

\emph{(a) If $g_v(\rho)\ne g_v(\rho')$ for some $\rho,\rho'$ that differ exactly in coordinate~$i$, then $c_v(B)=c_v(C)=i$.} Choose $y\in\Gamma_C$ whose level is the middle one of its cell on every axis $l\ne j$, and choose $x,x'\in\Gamma_B$ that differ in every coordinate and have patterns $\rho$ and $\rho'$ against~$y$: on axis $i$, they lie on opposite sides of~$y_i$; on an axis $l\notin\{i,j\}$, they use the two levels on the prescribed side of~$y_l$; and on axis~$j$, they use two levels of~$B_j$. Then $v$ ranks $y$ between $x$ and~$x'$. Suppose that $c=c_v(B)\ne i$, and let $\tilde x$ and $\tilde x'$ arise from $x$ and $x'$ by exchanging their $c$-th coordinates. They lie in~$\Gamma_B$, differ in every coordinate, and have the same patterns against $y$ as $x$ and~$x'$, so $v$ ranks $y$ between $\tilde x$ and $\tilde x'$ in the same orientation. However, $v$ follows coordinate $c$ in~$B$, so it prefers $x$ to $x'$ if and only if it prefers $\tilde x'$ to~$\tilde x$, a contradiction. Hence, $c_v(B)=i$. Exchanging the roles of the two boxes gives $c_v(C)=i$.

\emph{(b) For $i\ne j$, the voter $v$ with $c_v(B)=i$ satisfies $c_v(C)=i$ and $g_v(\rho)=\rho_i$.} Choose $\rho$ with $\rho_i=+$ and exactly $\lfloor d/2\rfloor$ plus signs, so that the corresponding sign vectors, which also have a plus sign on axis~$j$, have $\lfloor d/2\rfloor+1$ plus signs, and let $\rho'$ arise from $\rho$ by changing~$\rho_i$. For pairs with these patterns, the grid digraph contains $x\to y$ but not $x'\to y'$, so by \Cref{lem:switch} some voter $v$ has $g_v(\rho)=+$ and $g_v(\rho')=-$. By (a), $c_v(B)=c_v(C)=i$, and by Step~2, $v$ is the only voter with $c_v(B)=i$. By (a) again, $g_v$ does not depend on any coordinate other than~$i$, so $g_v(\rho)=\rho_i$.

Since $v\mapsto c_v(C)$ is a bijection, (b) also gives $c_v(C)=j$ for the voter $v$ with $c_v(B)=j$. So $c_v(B)=c_v(C)$ for all voters~$v$.

\emph{Step 4: numbering the voters.} Any two boxes are connected by a sequence of boxes in which consecutive ones differ in one axis. So $c_v=c_v(B)$ does not depend on~$B$, and we let $v_i$ be the voter with $c_{v_i}=i$. By Steps~2 and~3(b), if $x\in\Gamma_B$ and $y\in\Gamma_C$ differ in every coordinate, and $B=C$ or $B$ and $C$ differ only in an axis other than~$i$, then $v_i$ prefers $x$ to $y$ if and only if $x_i<y_i$.

\emph{Step 5: marked points.} Fix $p\in P$ and~$i$. By (G2), voter $v_i$ prefers either all of $\Gamma_B$ to~$p$ or $p$ to all of~$\Gamma_B$. This side is the same for boxes $B$ and $C$ that differ only in an axis $j\ne i$: otherwise, say $\Gamma_B\Ri{v_i}p\Ri{v_i}\Gamma_C$, pick $x\in\Gamma_B$ and $y\in\Gamma_C$ that differ in every coordinate and satisfy $y_i<x_i$; by Step~4, $y\Ri{v_i}x$, a contradiction. So the side depends only on the cell~$B_i$.

Let $J$ be a cell of axis $i$ below $p_i$ and $J'$ one above~$p_i$. Choose cells of the other $d-1$ axes such that exactly $\lfloor d/2\rfloor$ of them lie below the corresponding coordinate of~$p$, and let $B$ and $B'$ be the boxes with these cells and $J$ or~$J'$, respectively. For $x\in\Gamma_B$ and $x'\in\Gamma_{B'}$, the grid digraph contains $x\to p$ but not $x'\to p$, so by \Cref{lem:switch} some voter prefers $x$ to $p$ and $p$ to~$x'$. A voter $v_l$ with $l\ne i$ ranks $x$ and $x'$ on the same side of~$p$ since $B$ and $B'$ have the same cell on axis~$l$. So it is $v_i$ that prefers $\Gamma_B$ to $p$ and $p$ to~$\Gamma_{B'}$. Hence, $v_i$ prefers an auxiliary point to $p$ if and only if its $i$-th coordinate is below~$p_i$.

Finally, let $p,q\in P$ with $p_i<q_i$, and let $x$ be an auxiliary point whose $i$-th coordinate lies in the cell directly above~$p_i$, which is below~$q_i$. Then $v_i$ prefers $p$ to $x$ and $x$ to~$q$.
\end{proof}

\subsection{The Reduction} \label{app:reduction}

We reduce $3$-inducibility of tournaments to $k$-inducibility for every $k\ge4$. The intended profile consists of a $3$-voter profile that induces the given tournament together with $k-3$ voters whose preferences over its alternatives are a fixed linear order or its reverse. For odd~$k$, these voters cancel out. For even~$k$, one of them remains, and we need the following lemma to know which pairs it turns into ties.

\begin{lemma}\label{lem:acyclic}
In every tournament, the arcs that lie on no directed triangle form an acyclic digraph.
\end{lemma}

\begin{proof}
Suppose that $y_1\to y_2\to\dots\to y_r\to y_1$ is a directed cycle of such arcs; then $r\ge4$. We show $y_r\to y_i$ for $i=1,\dots,r-1$ by induction on~$i$, which contradicts $y_{r-1}\to y_r$ for $i=r-1$. For $i=1$, this is an arc of the cycle. If $y_r\to y_i$ and $i\le r-2$, then $y_{i+1}\to y_r$ would close the directed triangle $y_r\to y_i\to y_{i+1}\to y_r$, which contains the arc $y_i\to y_{i+1}$. Hence, $y_r\to y_{i+1}$.
\end{proof}

Fix $k\ge4$, and let $T=(V,E)$ be a tournament with $n$ alternatives. By \Cref{lem:acyclic}, there is a bijection $\pi\colon V\to\{1,\dots,n\}$ with $\pi(x)<\pi(y)$ for every arc $x\to y$ of $T$ that lies on no directed triangle. Each alternative $x\in V$ becomes the point of $\mathbb Q^k$ with coordinates
\[
x_1=x_2=x_3=0,\qquad x_t=\pi(x)\ \text{for even }t\ge4,\qquad x_t=-\pi(x)\ \text{for odd }t\ge5.
\]
Among $4,\dots,k$, there are as many even as odd indices if $k$ is odd, and one more even index if $k$ is even. We add $2k$ \emph{reference points} $a_i$ and $b_i$ for $i=1,\dots,k$: $a_i$ has coordinate $n+2i$ on axis~$i$ and $n+2i+1$ on all other axes, and $b_i$ the other way round. So $a_i$ is below $b_i$ on axis~$i$ and above $b_i$ on all other axes. For every pair $\{x,y\}\subseteq V$, we add a grid $\Gamma^{xy}$ with $L(k,2k+2)$ levels per cell for the marked points $P_{xy}=\{x,y,a_1,b_1,\dots,a_k,b_k\}$, where $L$ is given by \Cref{lem:grid}. All levels are distinct from each other, across all grids, and from all coordinates of the other points.

The digraph $U$ has the alternatives in~$V$, the reference points, and all auxiliary points. Let $\lhd_i$ order them from the smallest to the largest $i$-th coordinate, where auxiliary points of the same grid with the same $i$-th coordinate are ordered by the numbering of the grid. The only pairs left unordered are pairs inside $V$ for $i\le3$. Pairs inside $V$ are oriented as in~$T$, except that, for even~$k$, the arcs of $T$ that disagree with $\pi$ are replaced by ties. All other pairs are oriented or tied as in the digraph induced by $(\lhd_1,\dots,\lhd_k)$. For odd~$k$, $U$ is a tournament. Every grid has a number of points that depends only on~$k$, so $U$ has $O(n^2)$ alternatives and can be constructed in polynomial time for fixed~$k$.

Since the original alternatives agree on the first three coordinates, \Cref{lem:grid} leaves the voters $v_1$, $v_2$, and~$v_3$ free on~$V$, and the reference points make the numbering of the voters the same for all grids.

\begin{lemma}\label{lem:padding}
For every $k\ge4$, the digraph $U$ is $k$-inducible if and only if $T$ is $3$-inducible.
\end{lemma}

\begin{proof}
Let $e=0$ if $k$ is odd and $e=1$ if $k$ is even. Suppose first that $(R_1,R_2,R_3)$ induces~$T$. For $i\le3$, let voter~$i$ rank the alternatives according to~$\lhd_i$ and rank~$V$, which forms an interval of~$\lhd_i$, according to~$R_i$. For $t\ge4$, let voter~$t$ rank the alternatives according to~$\lhd_t$, which ranks $V$ according to $\pi$ for even $t$ and in reverse for odd~$t$. On every pair that is not inside~$V$, each voter~$i$ agrees with~$\lhd_i$, so the pair is oriented or tied as in~$U$. Now let $x,y\in V$ with $\pi(x)<\pi(y)$. Voters $4,\dots,k$ contribute $e$ to the margin of $x$ over~$y$, and voters $1,2,3$ contribute $\mu_{(R_1,R_2,R_3)}(x,y)$. For odd~$k$, the pair is therefore oriented as in~$T$. For even~$k$, if $x\to y$ in~$T$, the margin is at least $1+1$, which gives the arc $x\to y$ of~$U$. If $y\to x$ in~$T$, this arc disagrees with $\pi$ and thus lies on a directed triangle of~$T$. By (D2), exactly one of the voters $1,2,3$ disagrees with it, so the margin is $-1+1=0$, which gives the tie of~$U$.

Conversely, suppose that a $k$-voter profile induces~$U$. For every pair $\{x,y\}\subseteq V$, the alternatives in $P_{xy}\cup\Gamma^{xy}$ induce a grid digraph, so \Cref{lem:grid} numbers the voters as $v^{xy}_1,\dots,v^{xy}_k$. Applied to the marked points $a_i$ and~$b_i$, the lemma shows that $v^{xy}_i$ is the only voter that prefers $a_i$ to~$b_i$. So $v_i=v^{xy}_i$ does not depend on $\{x,y\}$. Applied to $x$ and~$y$, the lemma shows that, for $t\ge4$, voter $v_t$ prefers $x$ to $y$ if and only if $x_t<y_t$. Hence, voters $v_4,\dots,v_k$ rank $V$ according to $\pi$ or its reverse and contribute $e$ to the margin of $x$ over $y$ whenever $\pi(x)<\pi(y)$. For odd~$k$, the margin of the voters $v_1,v_2,v_3$ on every pair inside $V$ thus equals the margin of all voters, so these three voters induce~$T$. For even~$k$, let $x,y\in V$ with $\pi(x)<\pi(y)$, and let $\mu$ be the margin of $x$ over $y$ of all voters. The margin of $v_1,v_2,v_3$ is $\mu-1$. If $x\to y$ in~$U$, then $\mu\ge2$ since $k$ is even, so $\mu-1\ge1$, and $T$ contains $x\to y$. Otherwise, the pair is tied in~$U$, so $\mu-1=-1$, and $T$ contains $y\to x$. In both cases, the voters $v_1,v_2,v_3$ agree with~$T$, so they induce~$T$.
\end{proof}

\begin{proof}[Proof of \Cref{thm:more-voters}]
Membership in NP is clear. For $k=3$, hardness is \Cref{thm:main}. For $k\ge4$, \Cref{lem:padding} reduces $3$-inducibility of tournaments, which is NP-hard by \Cref{thm:main}, to $k$-inducibility, and the constructed digraph is a tournament if $k$ is odd.
\end{proof}

The reduction is polynomial for every fixed~$k$, but $L(k,2k+2)$ stems from Ramsey's theorem for $2k$-element sets and is astronomically large. We use Ramsey's theorem because it gives a uniform argument for all~$k$. For specific values of~$k$, explicit gadgets lead to much smaller reductions; for $k=4$, for instance, a direct construction yields a digraph with $6n$ alternatives.

\end{document}